\documentclass{article}

\usepackage{arxiv}

\usepackage[utf8]{inputenc} 
\usepackage[T1]{fontenc}    
\usepackage{hyperref}       
\usepackage{url}            
\usepackage{booktabs}       
\usepackage{amsfonts}       
\usepackage{nicefrac}       
\usepackage{microtype}      
\usepackage{multirow}
\usepackage{lipsum}
\usepackage{graphicx}
\usepackage{amsmath}
\usepackage{nccmath}
\usepackage[super]{nth}
\usepackage{float}
\graphicspath{ {./images/} }
\usepackage{lscape}

\title{Smart Flood Resilience: Harnessing Community-Scale Big Data for Predictive Flood Risk Monitoring, Rapid Impact Assessment, and Situational Awareness}

\author{
    \sffamily\large \vspace{0.15in}
    Faxi Yuan\textsuperscript{1,*}, Chao Fan\textsuperscript{1}, Hamed Farahmand\textsuperscript{1}, Natalie Coleman\textsuperscript{1}, Amir Esmalian\textsuperscript{1}\\
    \sffamily\large \vspace{0.15in}
    \textbf{Cheng-Chun Lee\textsuperscript{1}}, \textbf{Flavia Patrascu\textsuperscript{1}}, \textbf{Cheng Zhang\textsuperscript{2}}, \textbf{Shangjia Dong\textsuperscript{3}}, \textbf{Ali Mostafavi \textsuperscript{1,*}}\\
    \sffamily\normalsize
    \textsuperscript{1}Zachry Department of Civil and Environmental Engineering\\
    \sffamily\normalsize
    Texas A\&M University, College Station, TX 77843, USA\\
    \sffamily\normalsize
    \textsuperscript{2}College of Technology, Purdue University Northwest, Hammond, IN 46323, USA\\
    \sffamily\normalsize
    \textsuperscript{3}Department of Civil and Environmental Engineering\\
    \sffamily\normalsize
    University of Delaware, Newark, DE 19716, USA\\
    \sffamily\normalsize
    \textsuperscript{*}Corresponding authors: faxi.yuan@tamu.edu, amostafavi@civil.tamu.edu\\
}

\begin{document}
\maketitle
\begin{abstract}
Smart resilience is the beneficial result of the collision course of the fields of data science and urban resilience to flooding. The objective of this study is to propose and demonstrate a smart flood resilience framework that leverages heterogeneous community-scale big data and infrastructure sensor data to enhance predictive risk monitoring and situational awareness. The smart flood resilience framework focuses on four core capabilities that could be augmented by the use of heterogeneous community-scale big data and analytics techniques: (1) predictive flood risk mapping; (2) automated rapid impact assessment; (3) predictive infrastructure failure prediction and monitoring; and (4) smart situational awareness capabilities. We demonstrate the components of these core capabilities of the smart flood resilience framework in the context of the 2017 Hurricane Harvey in Harris County, Texas. First, we demonstrate the use of flood sensors for the prediction of floodwater overflow in channel networks and inundation of co-located road networks. Second, we discuss the use of social media and machine learning techniques for assessing the impacts of floods on communities and sensing emotion signals to examine societal impacts. Third, we illustrate the use of high-resolution traffic data in network-theoretic models for nowcasting of flood propagation on road networks and the disrupted access to critical facilities, such as hospitals. Fourth, we leverage location-based and credit card transaction data in spatial analyses to proactively evaluate the recovery of communities and the impacts of floods on businesses. These analyses show that the significance of core capabilities of the smart flood resilience framework in helping emergency managers, city planners, public officials, responders, and volunteers to better cope with the impacts of catastrophic flooding events.
\end{abstract}

\keywords{smart resilience \and urban AI \and machine learning \and big data \and urban flood}

\section{Introduction}

Urban resilience, in the context of this paper, is the ability of systems—infrastructure, businesses, emergency response, and critical facilities—to maintain functionality needed for residents to reduce social, economic, physical, and well-being impacts of ﬂoods (Vugrin et al. 2010). The value of leveraging emerging datasets and analytics methods for augmenting disaster resilience capabilities is increasingly recognized among research and practice communities. For example, in the 2015–2030 Sendai Framework for Disaster Risk Reduction (UNDRR 2015), priority 1 emphasizes understanding disaster risk, which specifically acknowledges the value of emerging datasets from social media and mobile phone data for creating early warning systems, disaster risk modeling, assessment, mapping, and monitoring. A smart resilience approach is particularly advantageous in flood risk management and resilience. For instance, predictive flood monitoring and situational awareness are two core capabilities for the emerging field of smart resilience, whose goal is to augment community resilience with the use of smart technologies and systems (Anbarasan et al. 2020; Sun et al. 2020; Hughes et al. 2006). Predictive flood monitoring refers to the ability to anticipate imminent flood risks and impacts as an extreme weather event unfolds (Sood et al. 2018; Palen and Anderson 2016). Departing from the standard flood monitoring approaches using hydraulic and hydrological models that predict flood inundation levels for hazard mitigation and infrastructure improvements prior to flood events (Wang et al. 2019; Yin et al. 2016a, 2016b; Naulin et al. 2013), predictive flood monitoring focuses on prediction (e.g., next few hours) and nowcasting of spatial and temporal flood status based on the current status of flooding during the course of events. However, the current approaches for flood monitoring (Boukerche 2019; Silver and Matthews 2017; Alazawi et al. 2014) do not provide certain essential information (e.g., what areas will be inundated within the next few hours; what roads are likely to get inundated in the hours to come) to public officials, emergency managers, responders, and residents to better tailor their decisions and actions based on imminent events. While predictive flood monitoring evaluates near-future risks and infrastructure disruptions, situational awareness focuses on insights, such as assessment of population protective actions, flooding impacts, disrupted access to critical facilities, and spatio-temporal patterns of recovery. Situational awareness during response and recovery plays an important role in the ability of communities to cope with and reduce flood impacts (Huang and Xiao 2015; MacEachren et al. 2011; Vieweg et al. 2010). These insights are critical for public officials and emergency responders to effectively respond and to allocate resources during the response/recovery phase (Muhammad et al. 2020; Zhai et al. 2020).

The key to augmenting predictive flood risk monitoring, rapid impact assessment, and situational awareness is to harness community-scale big data (Praharaj et al. 2021; Meier 2015). Owing to the ubiquity of sensing technologies, community-scale big data, such as infrastructure sensor data, location-based population activity data, mobility data, and crowdsourced and social media data (Yabe et al. 2020; Schnebele et al. 2014), are available through “data for good” programs of commercial data aggregators and analytics companies (Neelam and Sood 2020; Ianuale et al. 2015). Community-scale big data enables capturing the dynamics of community status, flooding evolution, and the built environment (Eugene et al. 2021; Leitão et al. 2018). Following the timeline of a flood event, we propose to leverage heterogeneous datasets and data analytics to enhance predictive flood monitoring and situational awareness in a smart flood resilience framework (Cameron et al. 2012). Through a comprehensive review of existing studies, we identified the core capabilities needed in a smart flood resilience framework (figure 1) concentrating on predictive flood risk mapping and monitoring, rapid impact assessment, predictive infrastructure failure prediction and monitoring, and situational awareness capabilities (Anbarasan et al. 2020; Roy et al. 2020; Wing et al. 2020; Li et al. 2018; Cervone et al. 2016; Sun et al. 2020).

\begin{figure}[ht]
\centering
\includegraphics[width=\linewidth]{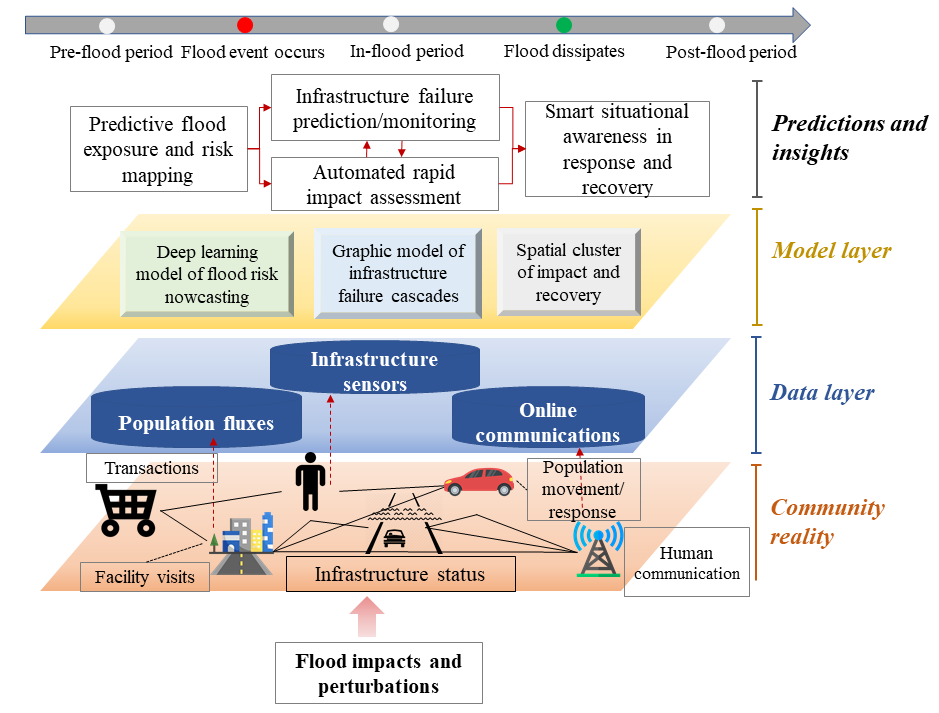}
\caption{Harnessing community-scale big data to capture the dynamic interactions among populations, infrastructure, and flooding to enhance fair predictive flood monitoring and situational awareness.}
\end{figure}

Using examples of studies in the context of the 2017 Hurricane Harvey in Harris County, Texas, this paper will further elaborate on achievement of the four capabilities in the smart flood resilience framework. First, for predictive flood risk mapping, we demonstrate the use of flood sensors for the near-future prediction of floodwater overflow in channel networks and the inundation of co-located road networks. Second, we discuss the use of social media and machine learning techniques for assessing the impacts of floods on communities and sensing emotion signals to examine societal impacts. Third, we illustrate the use of high-resolution traffic data in network-theoretic models for nowcasting of predictive infrastructure failure monitoring of flood propagation on road networks and the disrupted access to critical facilities such as hospitals. Fourth, we leverage location-based and credit card transaction data in spatial analyses to proactively evaluate the recovery of communities and impacts of floods on businesses. These examples demonstrate the significance of our proposed core capabilities of the smart flood resilience framework. These capabilities enable emergency managers, urban planners, and volunteers to better cope with the impacts of catastrophic flooding events. Finally, we discuss future research directions to advance the emerging and important field of smart resilience.

\section{Demonstration of smart flood resilience framework}

\subsection{Study context}
Harris County, Texas, is one the most flood-prone areas in the United States (figure 2). Hurricane Harvey made landfall in Texas Gulf Coast on August 25, 2017 and resulted in 120,000 flooded structures in Harris County alone (Harris County Flood Control District, 2021). During the four years since Hurricane Harvey, we have gathered datasets related to components of the smart flood resilience framework. The context of a large metropolitan city and the magnitude of Hurricane Harvey’s impact make Harris County a unique testbed for demonstrating how to harness community-scale big data through appropriate data analytic methods and machine learning techniques for implementing the smart flood resilience framework.

\begin{figure}[ht]
\centering
\includegraphics[width=0.7\linewidth]{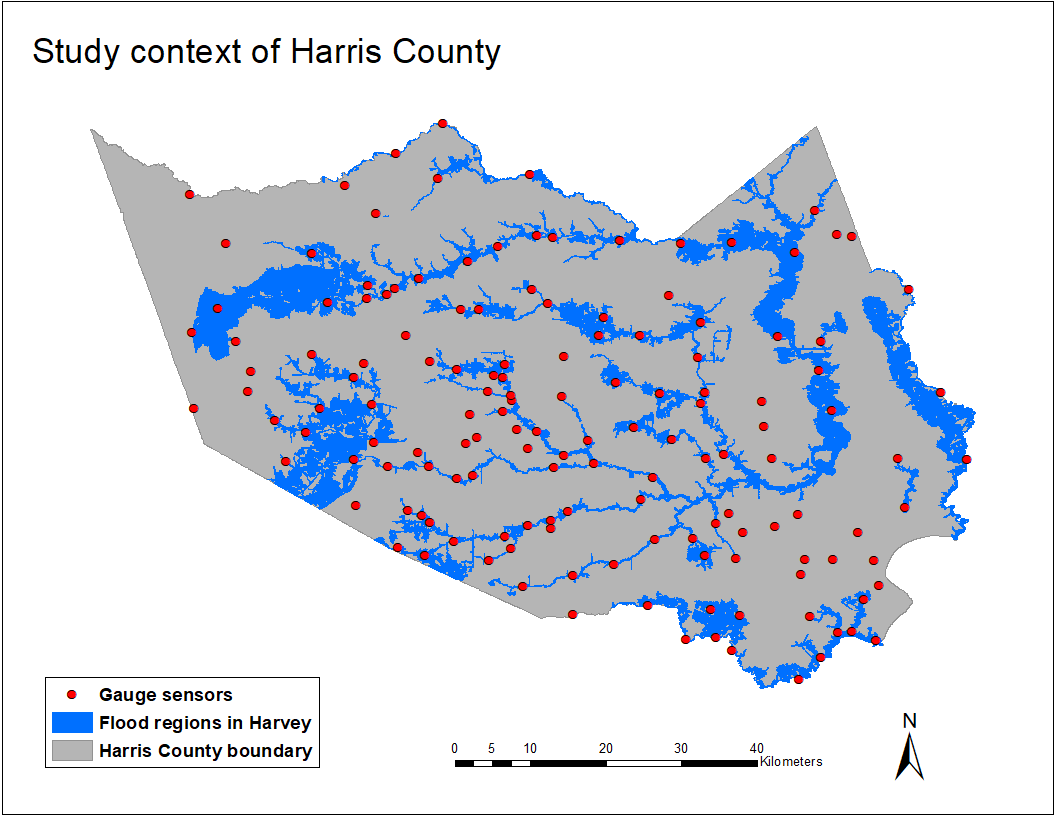}
\caption{Overview of Harris County. Blue regions represent flood zones of Harris County due to Hurricane Harvey. Red nodes refer to the locations of gauge sensor (for monitoring water depth) in Harris County.}
\end{figure}

\subsection{Overview of data and smart flood resilience capabilities}

In this section, we provide an overview of the four core capabilities (figure 3): (1) Predictive flood exposure and risk mapping capability focuses on using flood sensor data to predict failure probabilities of channels and roads which are co-located with channels through the use of machine learning techniques. The insights and foresight include which channels and roads (co-located with channels) are more likely to fail in the near future, which can inform emergency response efforts; (2) Automated rapid impact assessment capability focuses on utilizing social media data in machine learning techniques to monitor the unfolding of community disruptions and evaluate which communities are impacted; these insights can support situational awareness in response and recovery stages; (3) Predictive infrastructure failure monitoring is geared towards predicting the near-future road inundations and further evaluate which communities lose access to critical facilities (such as hospitals) due to road inundations. These insights are essential for situational awareness during the response and recovery; and (4) Situational awareness in response and recovery aims to monitor the spatial distribution and extent of impacts on business sectors and households and evaluate varying recovery durations across regions within a community. In the following sections, we will demonstrate the implementation of models related to each component. Examples of data used in each component are illustrated in Table 1. It should be noted that the capabilities related to different components of the smart flood resilience framework may be realized using different types of data and models. In this paper, we present examples of data and models for achieving these smart flood resilience capabilities.

\begin{figure}[ht]
\centering
\includegraphics[width=0.95\linewidth]{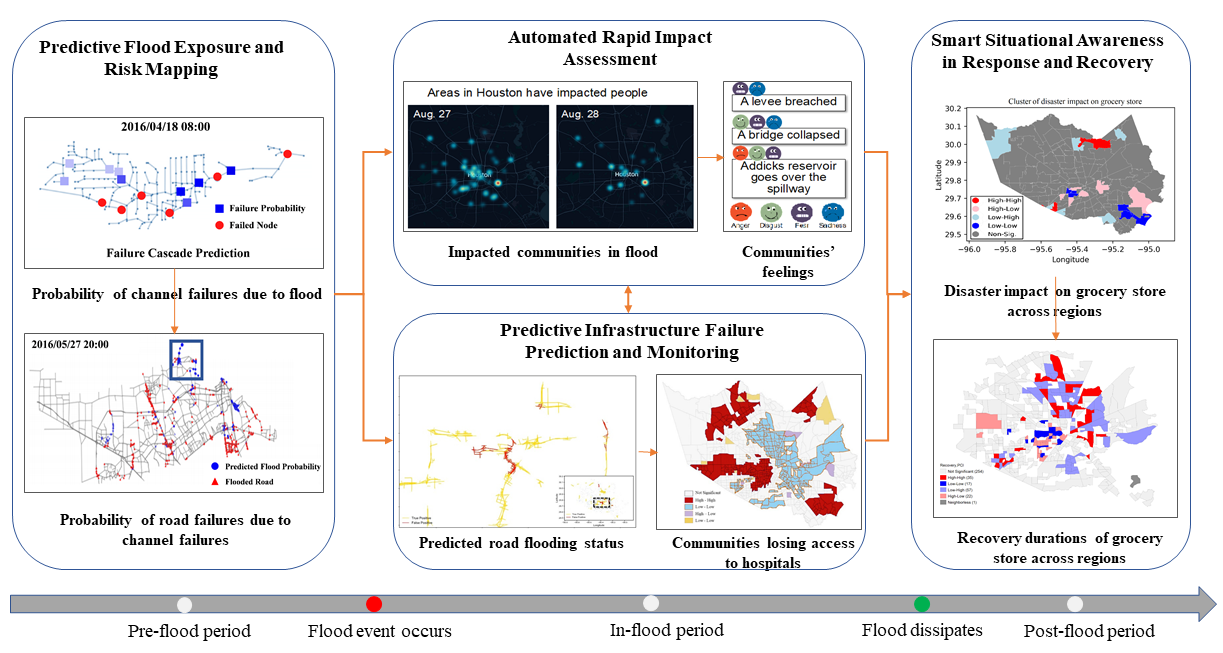}
\caption{Four core capabilities for smart flood resilience.}
\end{figure}

\begin{table}[]
\caption{Overview of data used in example studies}
\begin{tabular}{lll}
\hline
Dataset name                  & Data description                                                                                                                        & Data source/provider                                                           \\ \hline
Flood sensor data             & Dataset of flood gauge levels on channels in Harris County                                                                              & \begin{tabular}[c]{@{}l@{}}Harris County Flood\\ Control District\end{tabular} \\
Twitter data                  & $\sim$21 million tweets from users in Harris County                                                                                     & Twitter Gnip, Inc.                                                             \\
Road traffic data             & Acreage traffic speed on road segments in Harris County                                                                                 & IRINX, Inc.                                                                    \\
Credit card transaction data  & \begin{tabular}[c]{@{}l@{}}Number and total spend of transactions by ZIP code and \\ merchant category codes\end{tabular}               & Safegraph, Inc.                                                                \\
Points-of-interest visit data & \begin{tabular}[c]{@{}l@{}}Unique visit instances to physical locations in Harris\\ County from anonymized cell phone data\end{tabular} & Safegraph, Inc.                                                                \\ \hline
\end{tabular}
\end{table}

\section{Predictive flood exposure and risk mapping}

Predictive flood risk mapping identifies flood channels at risk of overflow and roads adjacent to channels likely to be inundated. Machine learning techniques are applied to infrastructure sensor data are to facilitate the near-future predictive flood risk mapping capability. The insights from this capability could support both rapid impact assessment and predictive infrastructure failures by emphasizing regions with high probability of getting flooded.

\subsection{Identifying channels at risk of failure and susceptible to failure cascade}
Channel networks are integral components of flood risk management in urban regions; however, discrete elements of a channel network may have varying probabilities of overflow. The failure of flood channels can, in turn cause flood propagation in surrounding neighborhoods. Data-driven techniques could facilitate the near-future predictive flood risk mapping of channels mainly by considering the spatial dependency of channels. Here, we demonstrate a data-driven Bayesian network model that integrates the topological structure of channel networks and historical hydrological data of flooding for predicting probabilities of flooding and failure cascade in channel networks. Hydrological data (water levels and rainfall intensity) collected from the flood gauges in the channel network were used to train and test the model. The channel network is abstracted as a graph where links represent the channels, and nodes represent the intersections of the channels. Failure probability for each channel is then calculated based on the trained data as well as the topology of the channel network. The Bayesian model allows calculation of the conditional probability of query variables with regard to the conditional probabilities and observed evidence. Target variables, which are the future status of inundation based on channel overflow, are inferred based on evidence variables: flood gauge measurement of rainfall through variable elimination operation. More details of the methods can be found in Dong et al. (2020a).

\begin{figure}[ht]
\centering
\includegraphics[width=0.65\linewidth]{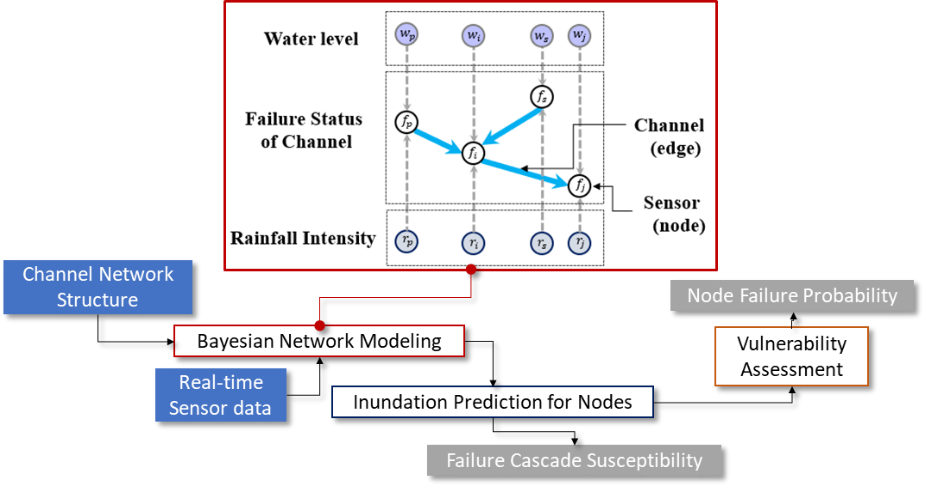}
\caption{The components and data flow of the Bayesian network modeling framework. The core Bayesian network structure is built upon the channel network structure. Two algorithms were developed to perform vulnerability assessment and inundation prediction for nodes based on the trained Bayesian model. The output of inundation prediction enables quantifying failure cascade susceptibility, and the vulnerability assessment enables determining node failure probability.}
\end{figure}

The data-driven Bayesian network model was trained and tested in Harris County using data from three historical flood events. The data related to Hurricane Harvey and the Memorial Day Flood (2015) were used for model training. Tax Day Flood (2016) data were used for testing based on rainfall intensity and water level in channels were extracted from the Harris County Flood Warning System. The model shows a capability for predicting the failure cascade in the channel network. Hence, the model provides valuable insights for predictive flood risk mapping. Figure 5 shows an example of the failure cascade captured by the model during the Tax Day Flood in one watershed in Harris County. We can see that at 2016-04-18 04:00, the predicted node with the highest failure probability (figure 5a), which is located in southwest of Brays Bayou, will be flooded in the next time step (figure 5b). Moreover, at 2016-04-18 08:00, three nodes on the main channel downstream of Bayou show a high probability of flooding in the next time step.

\begin{figure}[ht]
\centering
\includegraphics[width=0.85\linewidth]{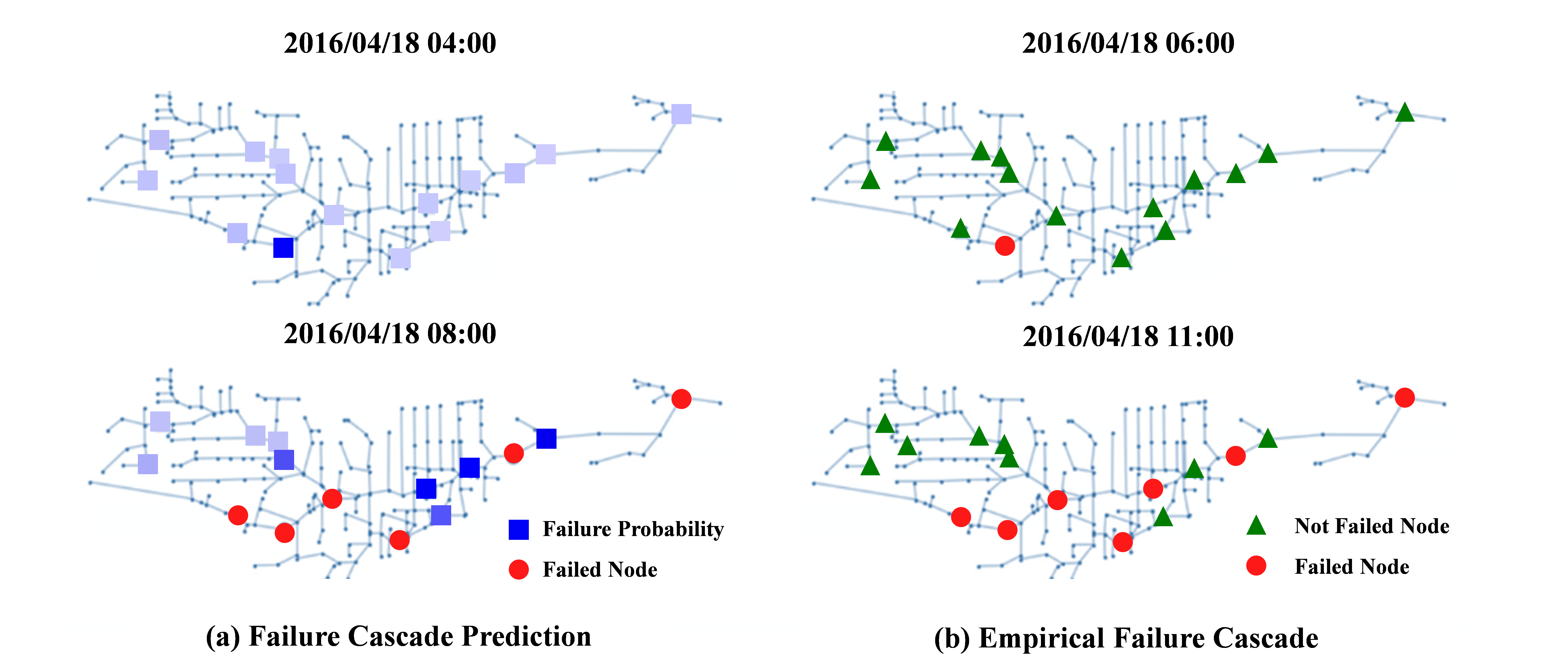}
\caption{An illustrative example of flood prediction for characterizing failure cascades versus the empirical failure cascade in Brays Bayou watershed; (a) results of simulation of the failure probability at each node (the darker blue represents the higher failure probability). (b) the empirical flood situation on in next time steps.}
\end{figure}

The predictive ability of the model helps flood monitoring for near-future risk mapping. In addition, the predictive monitoring capability of the model enhances situational awareness as a pivotal need for residents and decision-makers during a flood crisis. For example, residents living in neighborhoods near channels that are predicted to experience overflow could be evacuated hours before it happens. A quantitative measure capturing dependencies between channels susceptible to failure cascade (FCS measure) and spatial mapping of failure cascade to capture dependencies between channels identify channels with higher susceptibility to failure cascade, providing tools to prioritize in infrastructure project development and hazard mitigation plans. Neighborhoods with high FCS channels need to be prioritized in flood risk preparedness due to their susceptibility to cascading failure. Moreover, the FCS map (figure 6) reveals channels with high failure cascade susceptibility that need to be considered for proper flood monitoring.

\begin{figure}[ht]
\centering
\includegraphics[width=0.65\linewidth]{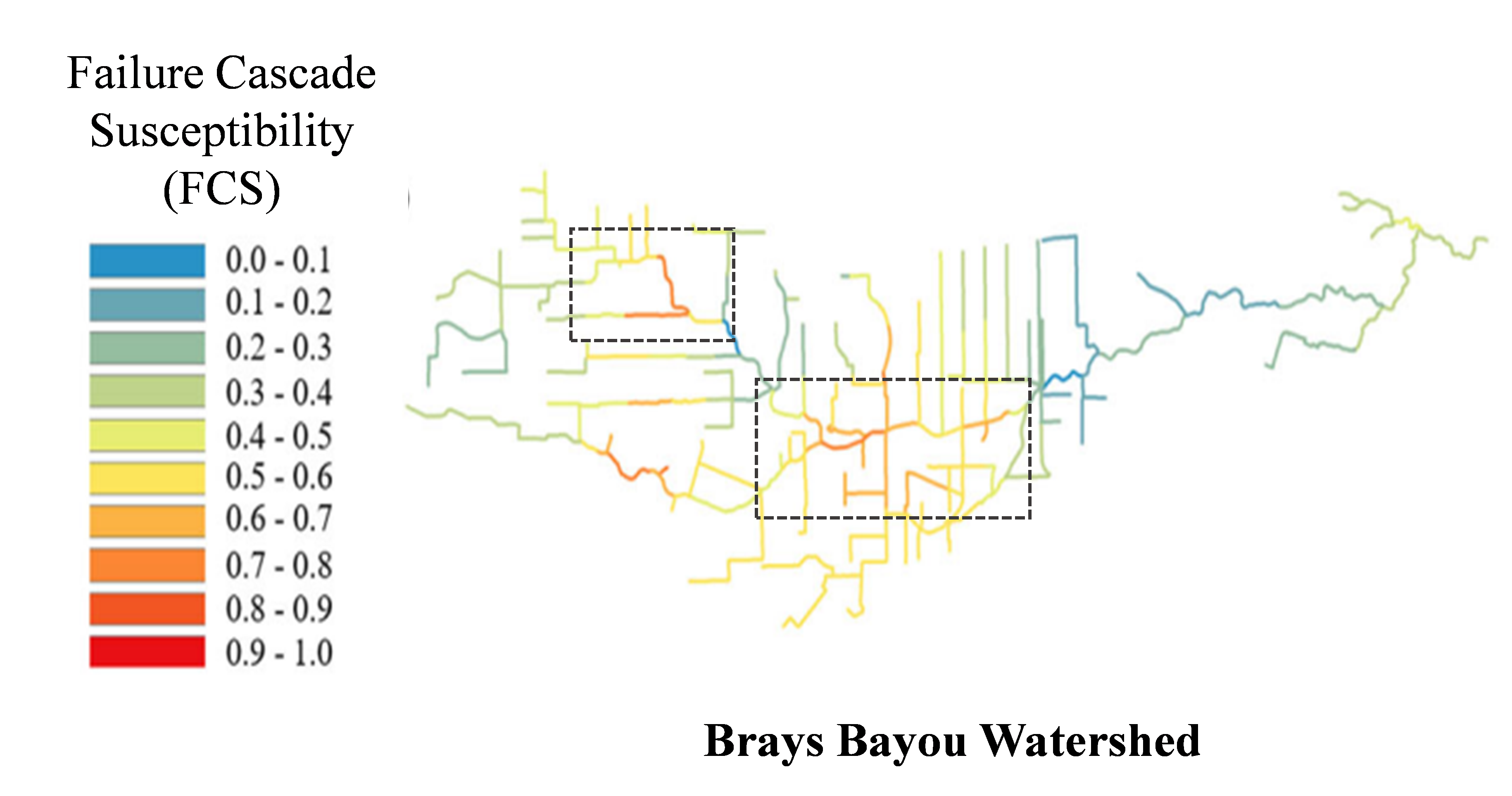}
\caption{Failure cascade susceptibility of flood control network in Brays Bayou watershed; we can observe that channels in the south central part and northeast part of the Bayou are prone to failure due to the influence of the failure of other nodes.}
\end{figure}

\subsection{Predicting roads susceptible to flood cascade failure from co-located channels}
As failure in one system is likely to propagate to its dependent counterpart and result in cascading failure, interdependency between infrastructure systems must be captured flood risk mapping. For co-located channel-road networks, when the water level in channels exceeds the bank, co-located roads can be inundated. In this section, a data-driven probabilistic graphical model is presented to predict road failure probabilities by capturing the co-location dependencies between channel and road networks. The model builds a Bayesian network structure by incorporating the topology of channel and road networks, along with their co-location dependencies. The proposed framework (figure 7) consists of three steps: (1) mapping the channel network and interpolating the flooding status of the intermediate channel components based on the existing sensors; (2) capturing and mapping the co-location dependency between the channel network and road network to construct the model structure; and (3) forecasting the probability of inundation in road segments inferred from the data collected from flood sensors that record water level and rainfall intensity of locations on the channel network. Using maximum likelihood estimation (MLE), the Bayesian network model calculates the difference between the flooding probabilities when its matched sensor is inundated and when its matched sensor is not inundated to estimate the flood risk of road intersection. Then, historical flood sensor data is collected and fed into the model to characterize failure cascades, evidenced by inundation of a road network. Similar to the model in sub-section 3.1, the model was trained and tested on historical flood events in the region. More information about this model can be referred to Dong et al. (2020b).

\begin{figure}[ht]
\centering
\includegraphics[width=0.75\linewidth]{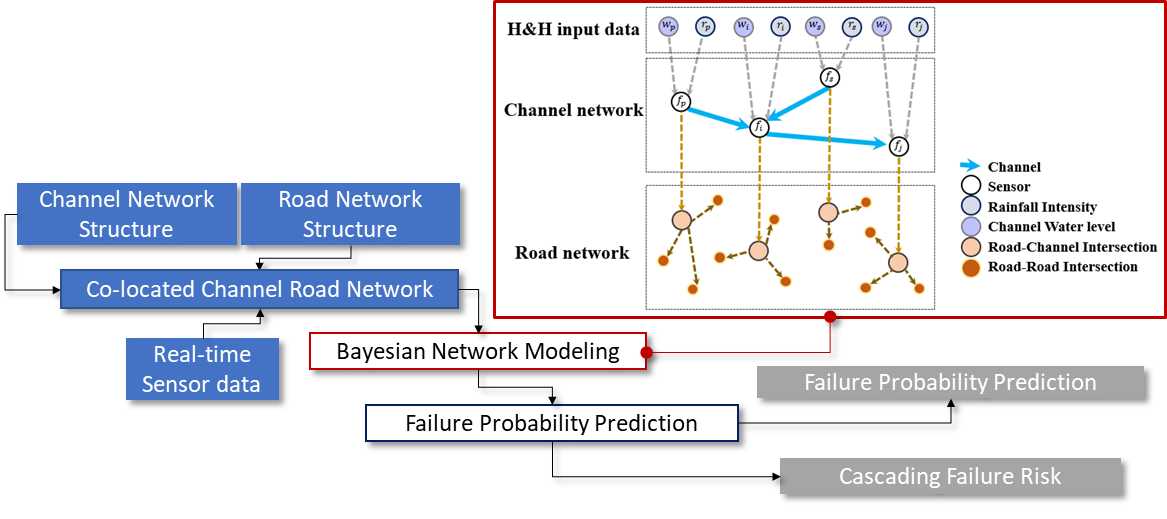}
\caption{The components and data flow of the co-located road-channel Bayesian network model. This model takes dynamic hydrological data collected from sensors as input. Failure probability is then calculated for inundation prediction on roads nearby channels.}
\end{figure}

Figure 8a shows the predicted flooding probability at each intersection of the road network. The shades of blue indicate the higher flooding probability in the next time step. (Only nodes with a probability large than 0.5 are plotted.) Figure 8b illustrates an example of the actual inundation map of the road intersections in a 4-hour interval. Green nodes indicate unflooded intersections. In both figures, red nodes represent the flooded road intersections. Blue boxes in figures 8a and 8b show that the flood propagates in road intersections with a high predicted failure in the following time step. The maps indicate the capability of the proposed data-driven model to predict the flood propagation dynamics in a co-located channel-road network. This early warning information can enable communities and crisis responders better respond to road inundations by avoiding roads with high flooding probability when choosing evacuation routes. In addition, the time series of the predicted failure probability can be extracted from the model during the course of the flood event. 

\begin{figure}[ht]
\centering
\includegraphics[width=0.85\linewidth]{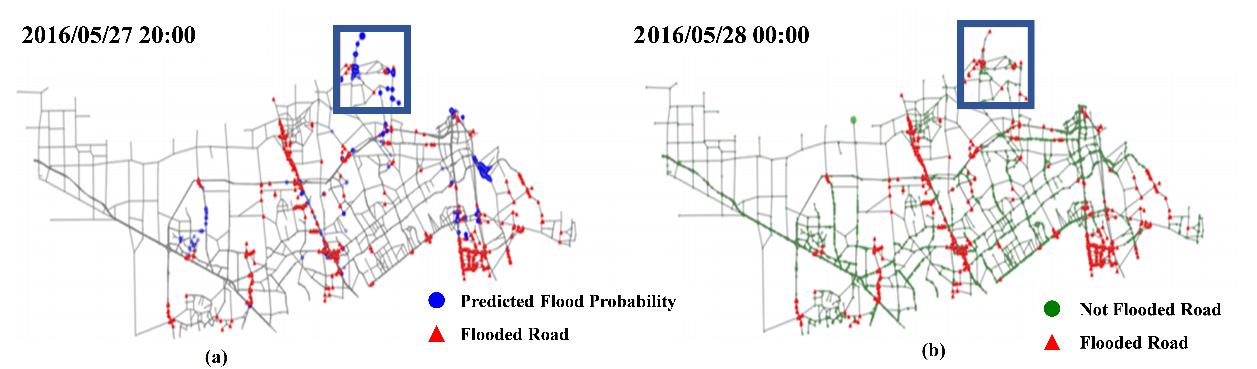}
\caption{The prediction capability of the model for detecting temporal road failure cascade; (a) the predicted flooding probability for road intersections, (b) the empirical flood situation in the next time step. Results show that in the North part of the Bayou, the roads with a high predicted flood probability are mostly flooded in the next time step.}
\end{figure}

Crisis responders can identify the areas of high flooding risk in near future using the plotted time series and the changes in flooding probabilities. Similarly, the model output helps quantify the risk of cascading failure from channel network to road network using a universal cascading risk index for each intersection (figure 9). A data-driven index can be obtained from the trained model to represent the overall cascading failure risk for roads, which encapsulates the risk of inundation propagation independent from the flood scenario. Calculating the cascading failure risk for each road, a failure cascade risk map can be plotted and used as a decision support tool for flood emergency response representing the magnitude of inundation risk in terms of failure cascade. The clusters of roads that are highly exposed to cascading failure due to the overflow in co-located channels can be identified and proper actions can be planned for avoiding considerable loss of functionality in the road network, especially if the roads with high failure cascade risk are located near the critical intersections of the network.

\begin{figure}[ht]
\centering
\includegraphics[width=0.85\linewidth]{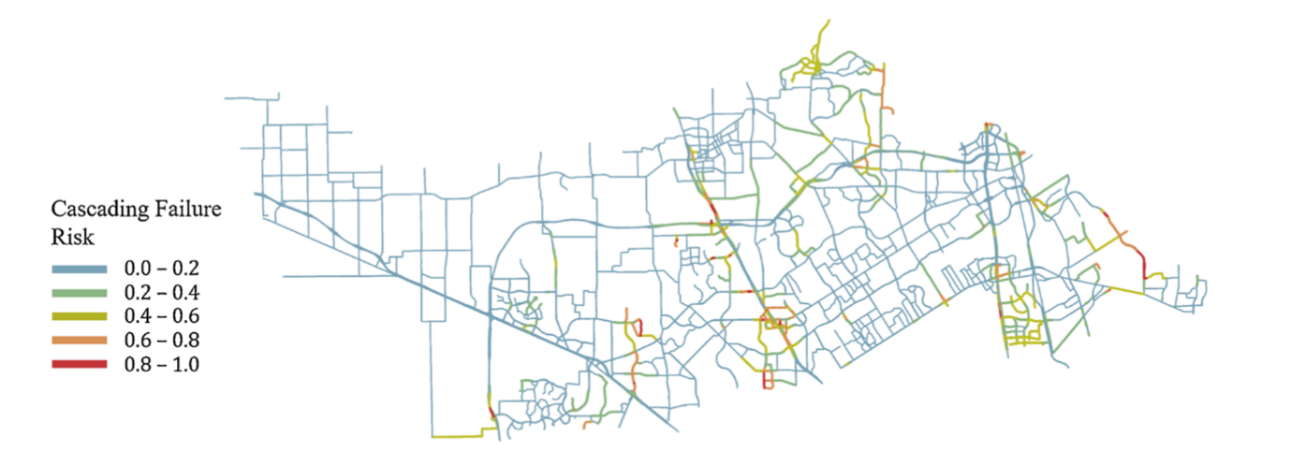}
\caption{Cascading failure risk index for Brays Bayou. Results indicates that cluster of roads areas southeast and south-center of the Bayou are highly influenced by the inundation of other roads.}
\end{figure}

\section{Automated rapid impact assessment}
Rapid impact assessment supports responders in formulating real-time response strategies, such as making informed decisions, better-allocating relief resources, and prioritizing site visits and response initiatives (Adam et al. 2012; Plank 2014; Cervone et al. 2016). Automated rapid impact assessment quickly evaluates the extent of flood impacts in a timely manner (Yuan and Liu 2018, 2020). Through two examples in this section will show how harnessing community-scale big data with machine learning approaches could enable rapid flood impact assessment. The first study investigates which infrastructure was disrupted and where it occurred during floods. The second study examines how people are feeling regarding community disruptions. Automated rapid impact assessment capability and the available community-scale big data are not limited to the examples and data categories discussed in this section.

\subsection{Social sensing of community disruptions and relief events}
Assessment of community disruptions in near real-time is necessary for effective responses in disasters. Prior studies have harnessed a vast array of technologies to process social media data to evaluate community disruptions (Fan et al. 2019). However, geotagged social media posts are rare and usually do not have sufficient and specific locations information to pinpoint the location of disruptive events. Recent advances in the detection of location and finer-grained event information could improve the utility, credibility, and interpretability of social media data for situation awareness (Fan et al. 2020a). The method, a hybrid machine learning pipeline, integrates named entity recognition for detecting locations mentioned in the posts, location fusion approach to extract coordinates of the locations and remove noise information, fine-tuned BERT (bidirectional encoder representations from transformers) model for classifying posts with humanitarian categories, and graph-based clustering to identify credible situational information (figure 10).

\begin{figure}[ht]
\centering
\includegraphics[width=0.65\linewidth]{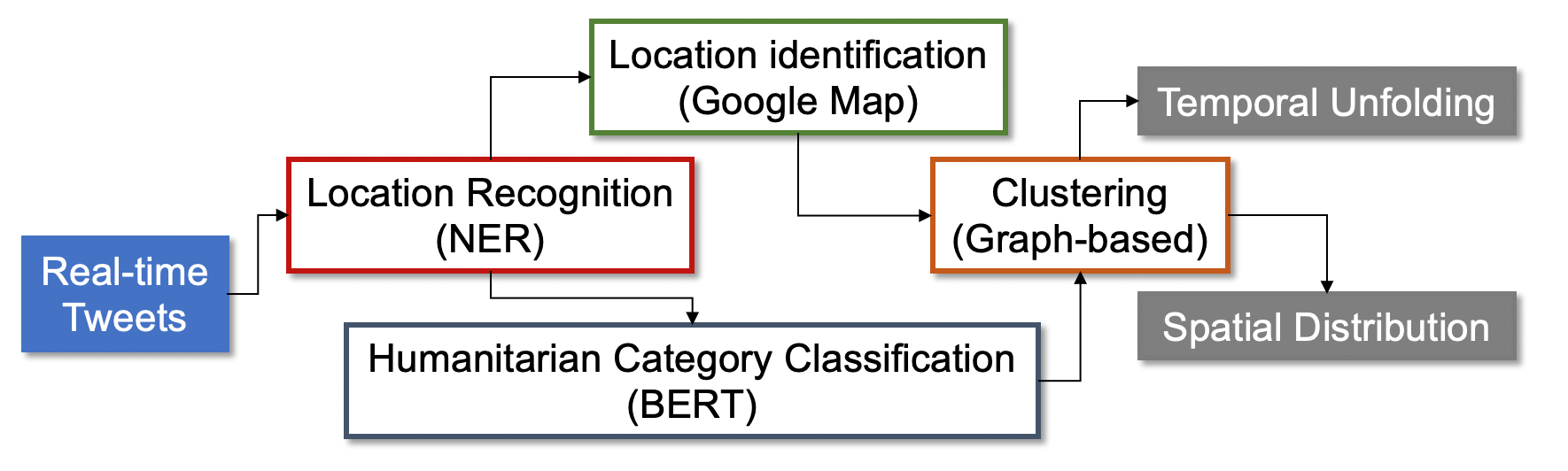}
\caption{The components and data flow of the hybrid machine learning pipeline. The model is composed of four computing components: location recognition using named entity recognition (NER), location identification using Google Map API, a BERT-based classifier for humanitarian categories of the tweets, and graph-based clustering to create semantic graphs for each category and location. The input would be real-time tweets, and the outputs are the timeline of events and the geographical maps showing the spatial distribution of the disruptive events.}
\end{figure}

Through the application of this method to the tweets posted during Hurricane Harvey in Houston, we can identify event locations using the location recognition algorithm and detailed information about the events using a graph-based clustering approach. For example, figure 11a shows the example of topic clusters for rescue, volunteering, and donation in the Houston Heights, and infrastructure and utility damage in Kingwood collected from Twitter data. The credible situational information delivered by the key tweets could show where the disruptions occur and how the events evolve. Furthermore, we summarized the situational information in a timeline to describe the temporal evolutions of the situation along with the dynamic disaster impacts. Figure 11b provides an example of the timeline of events related to rescue, volunteering, and donation in Houston. These time-tagged tweets mentioned where and what type of supplies or donations were provided and volunteer rosters. Finally, the method allowed capture of the locations to allocate the tweets collectively on a geographical map (figure 11c). In the map, we see the distribution of needs or donations reported on Twitter over time. This application shows the value of this method for automated mapping of events across fine-grained time and space using social media posts.

\begin{figure}[ht]
\centering
\includegraphics[width=0.85\linewidth]{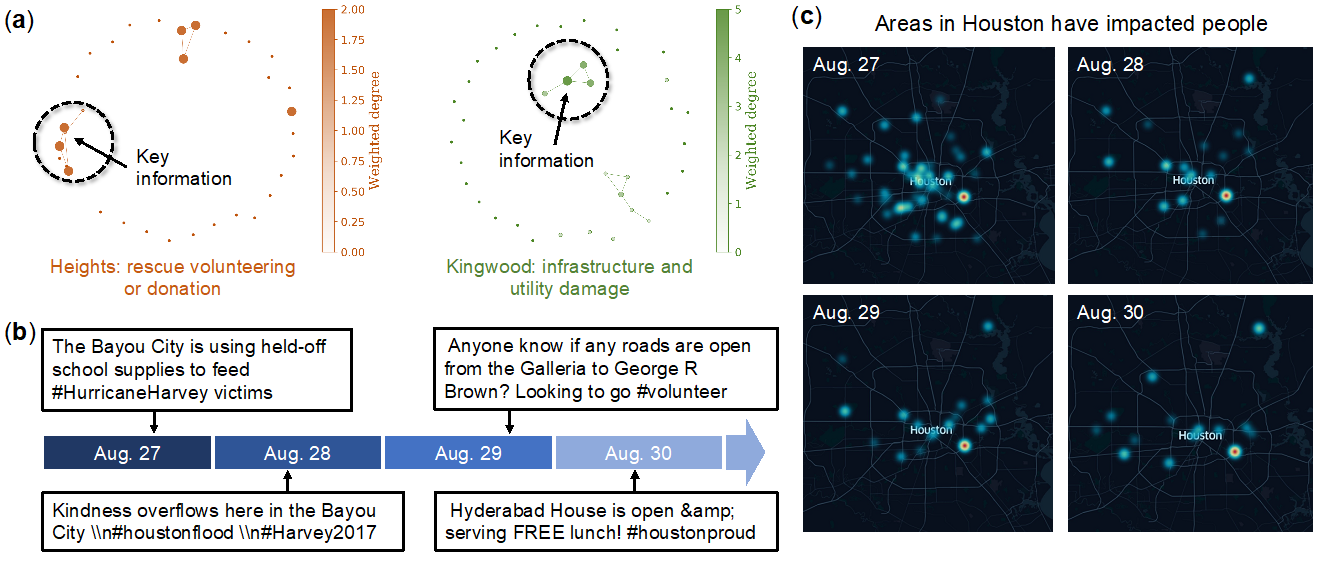}
\caption{(a) Identification of credible situational information from key tweets in topic-centric clusters: rescue, volunteering, or donation-related tweets posted in the Heights area; and infrastructure and utility damage in the Kingwood area. The edges in the figure are weighted based on the similarity scores. The color of the node represents the weighted degree of the node. (b) Timeline of events relevant to rescue and volunteering in Houston. The content in the boxes is the texts in the tweets including the main texts and the hashtags. (c) Mapping of areas in Houston with the density of reported people impacted by the event. The events are concentrated in the areas that are highlighted in the figures. The light shows the density of the events posted in tweets.}
\end{figure}

Using this method, crisis first responders can rapidly obtain situational information. The distribution of impact events provides a spatial view of the disaster impact on infrastructure and populations. Volunteers and relief organizations can identify areas in need and allocate their resources to locations where needs have not been satisfied. 

\subsection{Sensing emotion signals caused by community disruptions}
The analysis of aggregated, anonymized social media data serves a complement to information regarding built environment for purposes of assessing the status of the population during extreme weather events. Prior studies have explored and measured societal impacts of community disruptions using statistical data and surveys (Esmalian et al. 2019). While these collected data could further the understanding real-time effects of disasters, gathering and extracting insightful information is time-consuming. Publicly available social media content is a rich source information about the human condition and coping mechanisms during disruptions. To leverage social media analytics into the measurement of needs and reactions of the public, a social media analytics system, Social Sensing of Disaster Impacts and Societal Considerations (SocialDISC), has been proposed (Zhang et al. 2020). This approach (figure 12) comprises three main analyzers: a weakly supervised categorization model ensembled with a multichannel bidirectional long short-term memory (Bi-LSTM) classifier and fine-grained taxonomy; a knowledge distillation model which could detect sub-events related to disruptions; and emotion annotation and scoring method to identify the emotional signal from the tweets.

\begin{figure}[ht]
\centering
\includegraphics[width=0.85\linewidth]{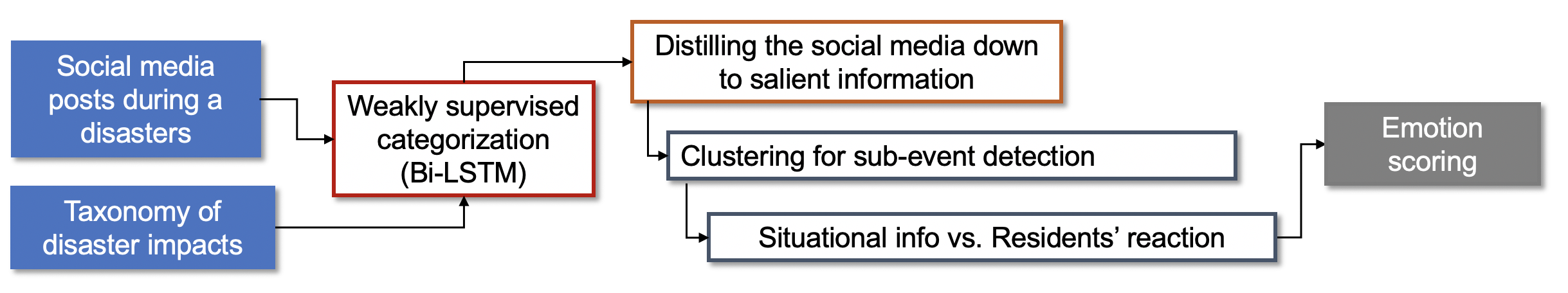}
\caption{The framework for the SocialDISC model for retrieving disruptions of critical infrastructures and societal impacts. The proposed model takes as input social media posts during a disaster fits them to a predefined taxonomy of disaster impacts. SocialDISC uses a weakly supervised approach to label social media posts with event categories. An artificial recurrent neural network architecture, Bi-LSTM, was trained. Then the content distilling process was accomplished with a network-based clustering algorithm to categorize social media posts into clusters according to content similarity. Finally, the approach detects and scores the emotion signal within residents’ reaction posts.}
\end{figure}

The application of this approach was conducted in the case of the 2017 Hurricane Harvey in Harris County. Using a great number of tweets posted by users living in Houston, we mapped how the status of infrastructure evolved and human emotions in reaction to infrastructure disruptions. The weakly-supervised approach with the Bi-LSTM classifier labels the tweets with specific event categories. Then the graph-based clustering algorithm detects sub-events from tweets in the same category. Figure 13 shows some examples of sub-events for flood control infrastructures. At the beginning of the disaster, reservoir water level increased and bayous overflowed. This eventuality was the major concern of the public, and was also the source of negative emotions, such as anger and disgust. As the hurricane evolved, flooding propagated and the decision was made to release water from Addicks and Barker retention reservoirs to prevent damage to their dams. People feared flooding in their communities and felt sad about breached levees and the collapsed bridge, which registered anger and disgust emotions. The findings allowed us to capture the impacts of community disruptions on residents’ emotions. 

\begin{figure}[ht]
\centering
\includegraphics[width=0.85\linewidth]{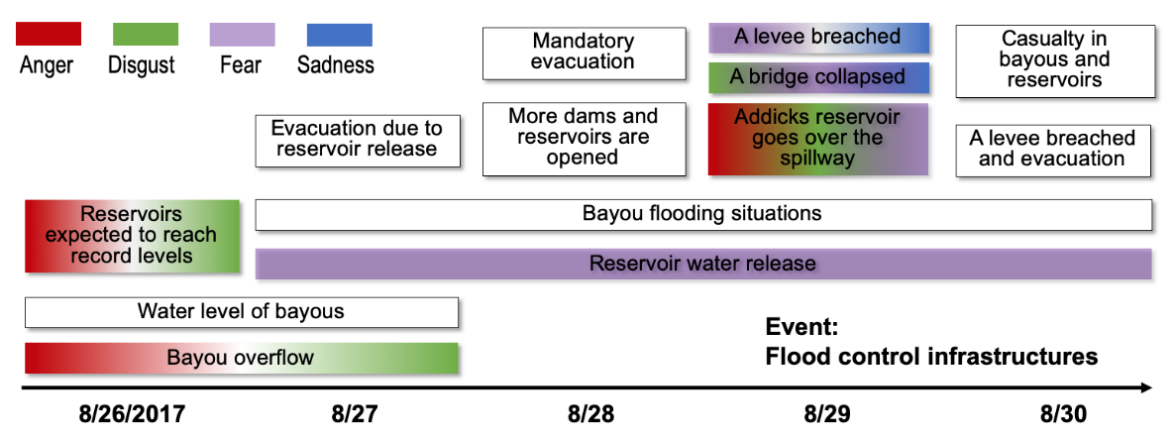}
\caption{Social impact analysis based on Twitter data. Colors represent the emotion signal detected from the content of the tweets, which are related to specific sub-events. For example, at the beginning of the hurricane, people were angered about rainfall and channel overflow, and they are concerned about the water release from reservoirs. When a levee breached and a bridge collapsed, people were sad and fearful.}
\end{figure}

The SocialDISC tool could support decision-makers to identify the gap between the public’s expectation for infrastructure systems and the capability of existing facilities in disasters. Through detected emotion signals related to the societal impacts of disruptions from the tweets, crisis responders can evaluate public depression and make intelligent preparation for depression mitigation. For example, people have varying tolerances for disruptions of different infrastructure and community services. The understanding of human emotions provides a metric for the people’s tolerance of disruptions and the evidence for response prioritization. In particular, the loss of food supplies may cause more negative emotions than that of power outage. Under the constraints of manpower and transportation capabilities, first crisis responders could prioritize resources accordingly.

\section{Predictive infrastructure failure prediction and monitoring}
Timely awareness of flood status of road networks enables emergency management agencies to understand which communities have lost access to essential facilities, such as hospitals and grocery stores (Yuan et al. 2021a) and informs residents of roads to avoid (Helderop and Grubesic 2019). In this section, we focus on two studies related to the road network flooding status and its further impacts on access to essential facilities such as hospitals (Yuan et al. 2021b; Fan et al. 2020b). These studies try to determine: 1) which roads will be inundated in a few hours 2) which neighborhoods will lose access to essential facilities due to road inundations. It should also be noted that predictive infrastructure failure monitoring capability includes many aspects, including communities without power and drinking water and is not limited to the examples discussed in this section.

\subsection{Nowcasting of flood propagation on road networks}
Near-future prediction (i.e., nowcasting) of road inundations is an aspect of situational awareness flood events focused on the near-future status. In this section, we present a mathematical model to predict flood propagation and recession within the first few hours as a flood event unfolds. The model consists of two components: flood contagion and network percolation process (figure 14). The flood contagion model, analogous to epidemiological models for pandemic prediction, is a mathematical model with a differential equation system. This component of the framework allows capture of the number of flooded road segments in the next 4 to 8 hours. The network percolation process could predict specific locations where the flooding would occur in the next couple of hours. The outcome of this work could identify specific locations of the flooded road segments in future hours.

\begin{figure}[ht]
\centering
\includegraphics[width=0.65\linewidth]{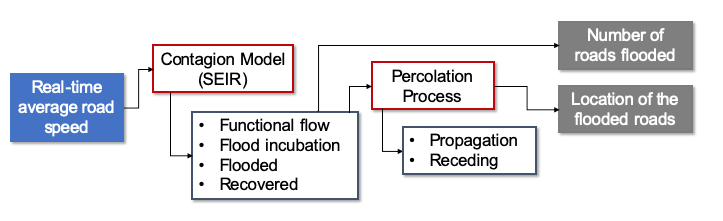}
\caption{The framework of the network percolation-based contagion model. The model takes as the input the real-time average road speed data, which can indicate the road flooding status. By knowing the number of roads that were flooded and not flooded, we run the SEIR (Susceptible-Exposed-Infected-Recovered) contagion model to predict the number of flooded roads in the next 4 or 8 hours. Then, the network percolation process is implemented to show the process of flood propagation and recession. The outputs of the model are the locations of the flooded roads in future hours.}
\end{figure}

The model has been applied to the case of the 2017 Hurricane Harvey in Houston to predict flooded road segments in the next few hours. Large-scale traffic data provided by INRIX, Inc., was used to estimate the flooded road segments (at each individual link at 15-minute intervals), serving as the ground truth for the predictive model. (INRIX collects anonymous traffic data from vehicles and provides data, analytics, and visualizations.) Figure 15 shows an example of predicted flooded road segments in the 4 hours following a specific timestamp during Hurricane Harvey. Based on the contagion model, which can accurately predict the number of flooded road segments, the network percolation process further identifies where the flooded road segments are located. Indicated by a large number of true positive road segments, the locations of flooded road segments in the next 4 hours are well-predicted. Only a small number of flooded road segments were not captured by the model. Through a more detailed view of the prediction results, the model identified the statuses of specific road segments, even in an area of highly dense road segments. The capability of this model would allow first crisis responders to estimate the scale and locations of flooding, evaluate the impacts of flooding on transportation systems, and further develop effective response strategies.

\begin{figure}[ht]
\centering
\includegraphics[width=0.7\linewidth]{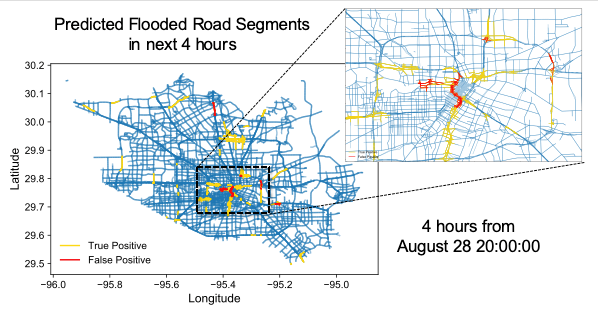}
\caption{Road flooding prediction using large scale traffic data. The predicted flooded road segments with both true positive and false positive results are shown in the figure. The model predicts the flooded roads in 4 hours from 8pm, August 28, 2017, in Harris County (Houston). The model shows good performance since the majority of the results are true positives. The inset shows details of the road segments and prediction results.}
\end{figure}

The outcomes of the model could provide valuable foresight to residents and emergency managers and responders. The model captures the spatial and temporal dynamics of flooding in urban road networks as well as the contagion effects of flooded road segments on their neighbors. This understanding can help us predict the time and locations of road inundations to proactively respond to flooding. Being aware of the flood situations in the coming hours, crisis responders could be informed about areas where evacuation would be difficult. Also, the insights from the model could help identify areas that could lose access to critical facilities (such as hospitals) as shown in the next section.

\subsection{Specifying neighborhoods most vulnerable to loss of access to hospitals}
Here, we present a data-driven model to integrate graph-based vulnerability assessments with the inputs that characterize the propagation of the flooding based on historical data to evaluate accessibility to critical facilities, in particular, hospitals. In this model, the road network is modeled as a graph, and hospitals are assigned to the closest roads. During the percolation simulation, roads are removed by a process in which removal probability is proportional to the inundation likelihood. More information about this model is available from Dong et al. (2020c). 

Based on the outcomes of the model, we can investigate whether there is a path between a neighborhood and any hospitals in the road network, and thus determine whether a neighborhood has access to hospitals. We implemented the framework in Harris County and simulated a flood scenario based on the characteristics of flooding caused by Hurricane Harvey. Figure 16 shows the components and data flow of the framework. The results of the probabilistic percolation simulation provide insights for crisis response and hazard mitigation. For example, clusters of neighborhoods with a higher risk of losing hospital access could be detected. Figure 17 shows clusters of census tracts with varying vulnerability in terms of losing access to hospitals during floods. The clusters are determined by using local indicators of spatial association (LISA) analysis and adopting Global Moran's I test to examine the similarity and dissimilarity of the neighboring census tracts in terms of vulnerability of access to hospitals. This result indicates large clusters of neighborhoods that are highly vulnerable to losing access to hospitals due either to a lack hospitals in proximity or an absence of reliable routes during Hurricane Harvey. Results also indicate that the neighborhoods vulnerable to losing access to hospitals are not necessarily located closer to floodplains and mainstream flows, specifically the areas downstream of Addicks and Barker reservoirs (box a in figure 17). A large cluster of census tracts with high vulnerability to losing hospital access is seen in boxes b and c (figure 17), while the extent of the area inside the floodplain is smaller. Other reasons, such as the low density of hospitals, vicinity of critical roads to channels with high inundation risk, or the topology of the road network can contribute to the vulnerability to loss of access. Moreover, in box b is a cluster of highly vulnerable census tracts that is similarly not extensively exposed to flood inundation based on the 100-year floodplain, which should be highlighted for hazard mitigation planning to ensure reliable access of residents to healthcare services during floods.

\begin{figure}[ht]
\centering
\includegraphics[width=0.65\linewidth]{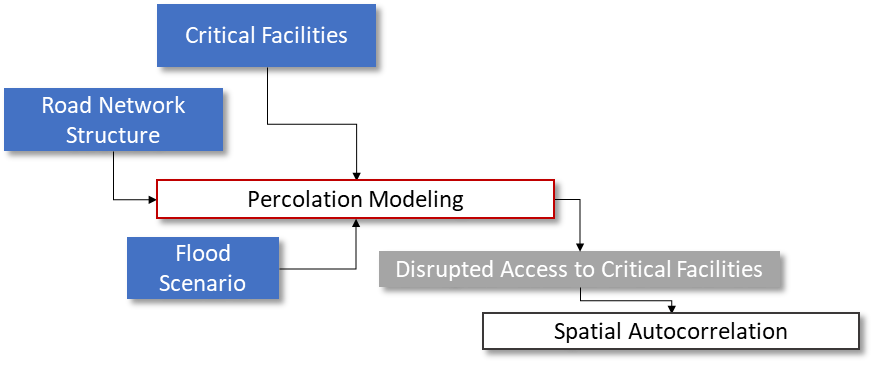}
\caption{The components and data flow of the analysis for detecting neighborhoods most vulnerable to loss of access to critical facilities. The model is composed of a percolation simulation process to identify areas with high vulnerability to loss of access to hospitals and a spatial autocorrelation that is used to assess the spatial distribution of vulnerable neighborhoods.}
\end{figure}

\begin{figure}[ht]
\centering
\includegraphics[width=0.65\linewidth]{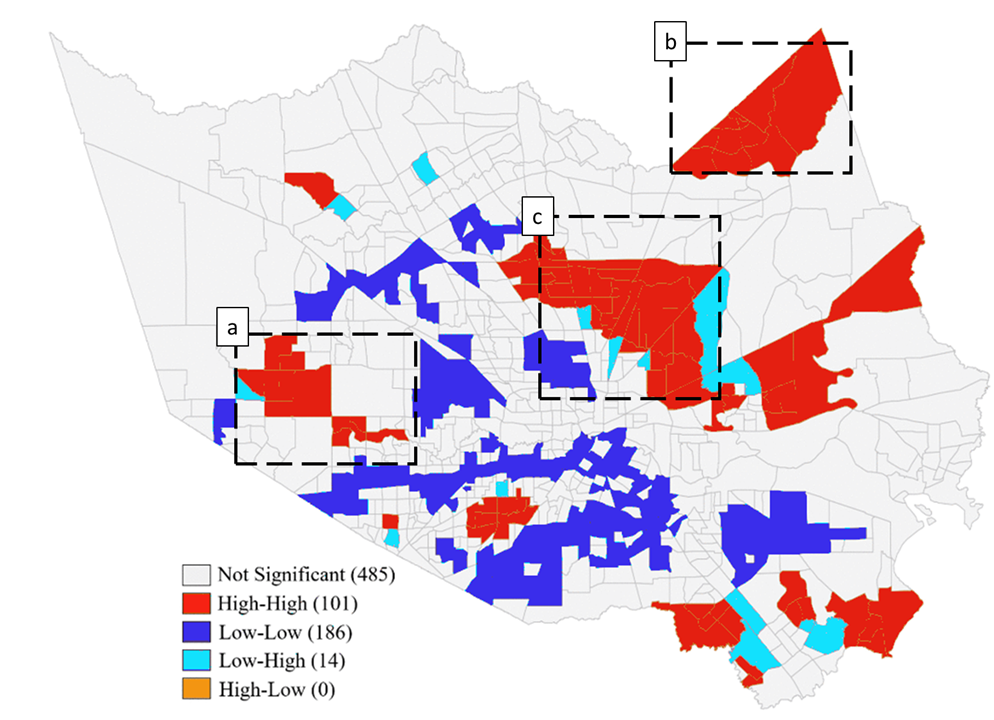}
\caption{Spatial distribution of vulnerability to loss of access to hospitals during Hurricane Harvey in Harris County. High-high reveals the clusters of regions with high vulnerability to lose access to hospitals surrounded by regions of high vulnerability; low-low denotes clusters of regions with low vulnerability surrounded by regions of low vulnerability. Low-high (high-low) reflects the cluster that regions with low (high) vulnerability are surrounded by those with high (low) vulnerability. Results show that there are large clusters of neighborhoods that are highly vulnerable to lose access to hospitals due either to lack of hospitals in proximity or absence of reliable routes to hospitals.}
\end{figure}

The method affords broad opportunities for practitioners in disaster management of infrastructure systems during floods. First, it provides insights beyond the topological analysis of road network vulnerability and assessment of the sufficiency of hospitals in neighborhoods. Moreover, results can help identify clusters of areas that have a higher risk of losing access to hospitals in a particular flood scenario. The map that shows the clusters of vulnerable areas can inform crisis responders to allocate sufficient resources to those areas to ensure that residents have access to healthcare services in case they lose access to hospitals. This map can also help residents in vulnerable neighborhoods understand risks they may encounter during extensive floods. Finally, urban development and transportation planners could consider the vulnerability maps for developing future projects by identifying neighborhoods that need to be prioritized for hosting healthcare infrastructures and/or require more redundancy in the road network to provide more reliable accessibility.

\section{Proactive monitoring of response and recovery}
To better support the response and recovery of flood-prone and affected areas, it is necessary to proactively monitor disaster impact and recovery based on the activities and attributes present in empirical data (Hikichi et al. 2017). Traditional data sources for disaster impact and recovery monitoring, such as surveys and interviews, are usually costly, time-consuming, and non-scalable (Boon 2014). With the emergence of new data collection technologies, community-scale data including visits to points of interest (POI) (Podesta et al. 2021), credit card transactions (Yuan et al. 2021c), social media data (Yuan et al. 2021d), crowdsourcing map data (See 2019), and mobility data from cellphones (Lu et al. 2016), have proven to be vital in timely situation monitoring of disaster impact and recovery during urban flooding events. In this section, we demonstrate two examples using POI visits and credit card transactions data to monitor the spatial and temporal patterns of flooding impact and communities’ recovery activities during Hurricane Harvey. Specifically, these two examples help to derive two essential insights for situational awareness: 1) which regions and which business sectors have suffered more severe disaster impact? 2) where and what essential facilities have experienced longer recovery duration? As noted in sections 3, 4 and 5, smart situational awareness capability includes many aspects (e.g., population displacement (Yabe et al. 2020)) and can be enhanced by community-scale big data; therefore, smart situational awareness is not limited to the examples and data categories to be discussed in this section.

\subsection{Evaluating which business sectors suffered severe impacts }
This section presents a study related to spatial patterns of disaster impact of business sectors based on analyzing fluctuations in credit card transactions (CCTs). Such fluctuations could capture the collective effects of household impacts, disrupted accesses, and business closures, and thus provide an integrative measure for examining disaster impact on business sectors across regions. This study used the fluctuations of CCTs to adapt the resilience curve and further derived the insight of disaster impact on business sectors across regions. The analytical framework was illustrated in figure 18. 

\begin{figure}[ht]
\centering
\includegraphics[width=0.7\linewidth]{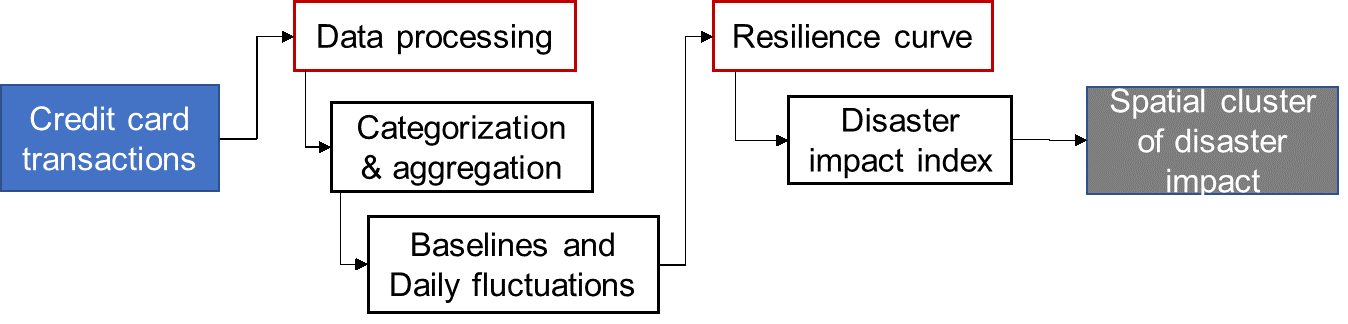}
\caption{Methodology for identifying spatial patterns of disaster impacts based changes in households’ credit card expenditures. Data processing includes credit card transaction categorization by merchant codes and aggregations by dates and regions (ZIP code). CCTs baselines were established on the CCTs in the first three weeks before Harvey, and daily fluctuations of CCTs were computed as the percentage of daily CCTs compared with the baselines. Thereafter, the resilience curve was established with daily CCT fluctuations (Nan and Sansavini 2017). The maximum drop of fluctuations of credit card transactions is defined as the disaster impact (a negative value). Then we used the local spatial autocorrelation method (local Moran’s I) to identify spatial clusters of disaster impact on business sectors. More detail of the method is in Yuan et al. (2021c).}
\end{figure}

With local Moran’s I for different ZIP codes, we identified the spatial clusters of disaster impact by LISA maps. The LISA cluster maps of ZIP codes can help identify regions with more severe disaster impact on business sectors which play a critical role in disaster preparedness, response, and recovery, such as drugstores and grocery stores (Beatty et al. 2019) (figure 19). Timely awareness of disaster impacts on these critical business sectors, such as drugstore and grocery stores, can help crisis responders identify regions with urgent needs of medical resources and grocery goods.

\begin{figure}[ht]
\centering
\includegraphics[width=0.85\linewidth]{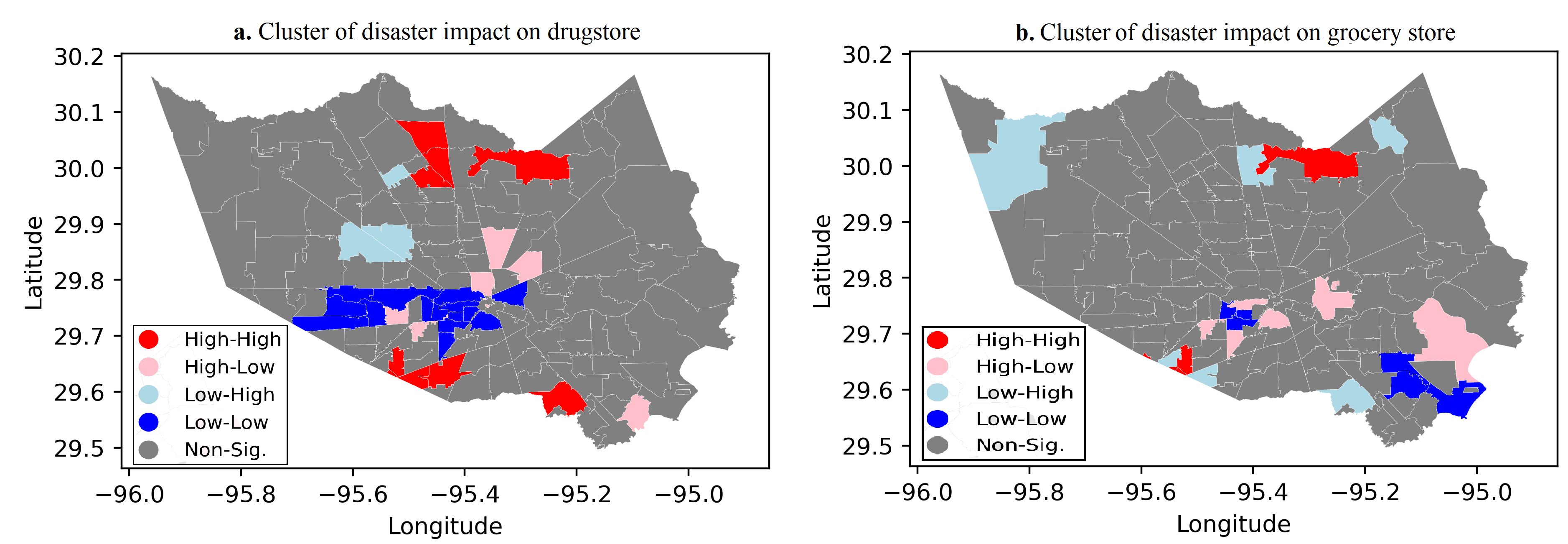}
\caption{LISA cluster maps for disaster impact on drugstore (a) and grocery store (b) sectors. In both figure 19a and 19b, blue clusters in represent regions with more severe disaster impact while red regions have suffered light disaster impact bases on credit card transactions in drugstores and grocery stores, respectively. High-high (Low-low) represents the clusters of ZIP codes with slight (severe) disaster impact. High-low (low-high) refers to the cluster pattern where ZIP codes with slight (severe) disaster impact were surrounded by those with severe (slight) disaster impact.}
\end{figure}

In figure 19, the ZIP codes shaded blue and red are clusters of similar values (high-high and low-low for disaster impact); the pink- and light blue-shaded areas are clusters with dissimilar values (high-low and low-high); grey-shaded ZIP codes do not present statistical significance for the spatial clusters (p-value > 0.1). According to figure 19, we can localize the ZIP code regions within the spatial clusters of more severe disaster impact (i.e., blue-shaded ZIP codes) on drugstore (a) and grocery store (b) sectors. Compared with the traditional survey methods that requires extensive time and resources and thus may fail to provide timely information about the complex interactions underlying disaster impact and recovery, harnessing CCTs provide rapid insight of disaster impact on these critical business sectors during the event. These rapid insights can guide public officials and crisis managers to deploy effective response and recovery strategies. For instance, figure 19 illustrates the ZIP code regions with severe disaster impact on drugstore and grocery store sectors (blue-shaded regions). Crisis responders can prioritize the delivery of medical resources and food and water support to these hotspots. In addition, integrating census data, such as population, with these hotspots, particularly the proportion of the elderly and low-income groups, crisis responders can appropriately deploy relief resources.

\subsection{Proactive monitoring of community recovery based on visits to essential facilities}
The majority of studies on community recovery from disasters have relied on surveys (Morss et al. 2016; Sherrieb et al. 2010). However, these studies are limited in their ability to measure and appropriately quantify recovery across fine spatial-temporal scales. Location-based data can characterize and provide insights on the spatial-temporal distribution of disaster impact by collecting information on population movements at different disaster stages (time) and at fine-grained areas (space). Figure 20 summarizes the conceptual methodology to examine the changes in time and distribution in space of points-of-interest (POIs), such as grocery stores, gasoline stations, and health and personal care facilities. Recovery duration refers to the period from the initial decrease point of POI visits to the eventual increase of POI visits to at least baseline levels.

\begin{figure}[ht]
\centering
\includegraphics[width=0.85\linewidth]{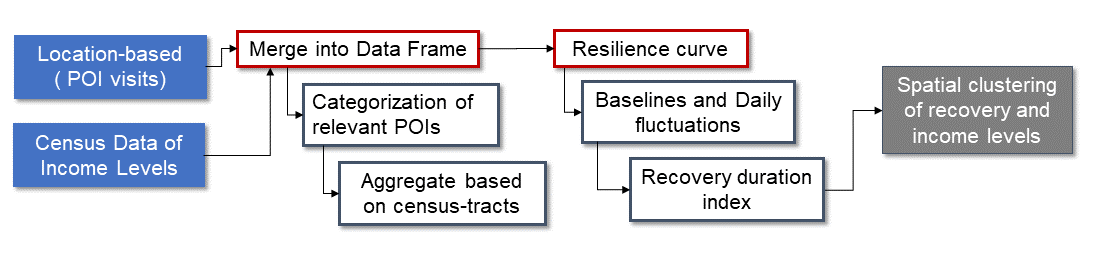}
\caption{Framework for processing and analyzing the change of time and distribution of space of location-based data. The location-based data (POI visits) was categorized into relevant categories, such as grocery store and gasoline stations. It was then merged with median income at the census-tract level. The baseline is determined by the daily average of the first three weeks of August, and then the noise is further filtered by applying a 7-day rolling average. Then, daily changing visits to POIs were computed by comparing daily POIs visits and their corresponding baselines. The recovery duration index shows the number of days from the date of landfall until a POI recovered from disruptions in visits to return to baseline levels. More details on the methodology can be found in Podesta et al. (2021).}
\end{figure}

In figure 21, the resilience curves are examples of the changes of POIs visits over time. Findings revealed that visits to certain POIs, such as grocery stores and gasoline stations, recovered more quickly than other POIs such as commercial stores and entertainment venues. This information could assist emergency planners in understanding which POIs in the community are most vulnerable to the disaster impact. It could also assist managers of POIs, such as business owners, determine the level of disruption after an extreme event.

\begin{figure}[ht]
\centering
\includegraphics[width=0.7\linewidth]{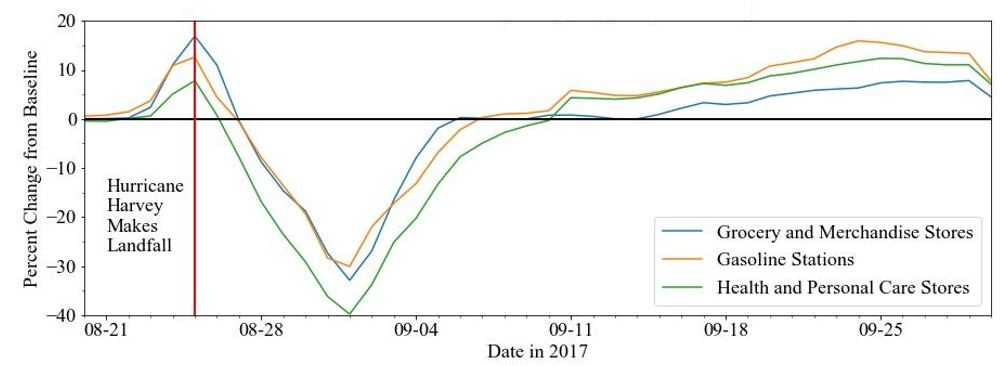}
\caption{Resilience curves capturing the change of visits from the baseline to the grocery and merchandise stores, gasoline stations, and health and personal care stores.}
\end{figure}

In addition, we explored how physical location impacted recovery duration. We merged sociodemographic information obtained from the US Census to the duration of recovery for the POIs for each census tract. The LISA cluster map of recovery duration and social vulnerability index (figure 22) visualized the extent of clustering of the duration of recovery for grocery stores to the per capita income of the census tracts. Such visual representations can assist emergency planners and other stakeholders in seeing where grocery stores had longer recovery duration and how different sociodemographic characteristics can influence the visits to POIs. The research can examine whether POIs have longer periods of recovery in spatially concentrated areas of higher social vulnerability such as low-income and minority households. Planners could assist low-income communities which experience longer durations of recovery by creating more economic opportunities and accommodate minority communities who experience longer durations of recovery by considering their distinct cultural and linguistic differences.

\begin{figure}[ht]
\centering
\includegraphics[width=0.65\linewidth]{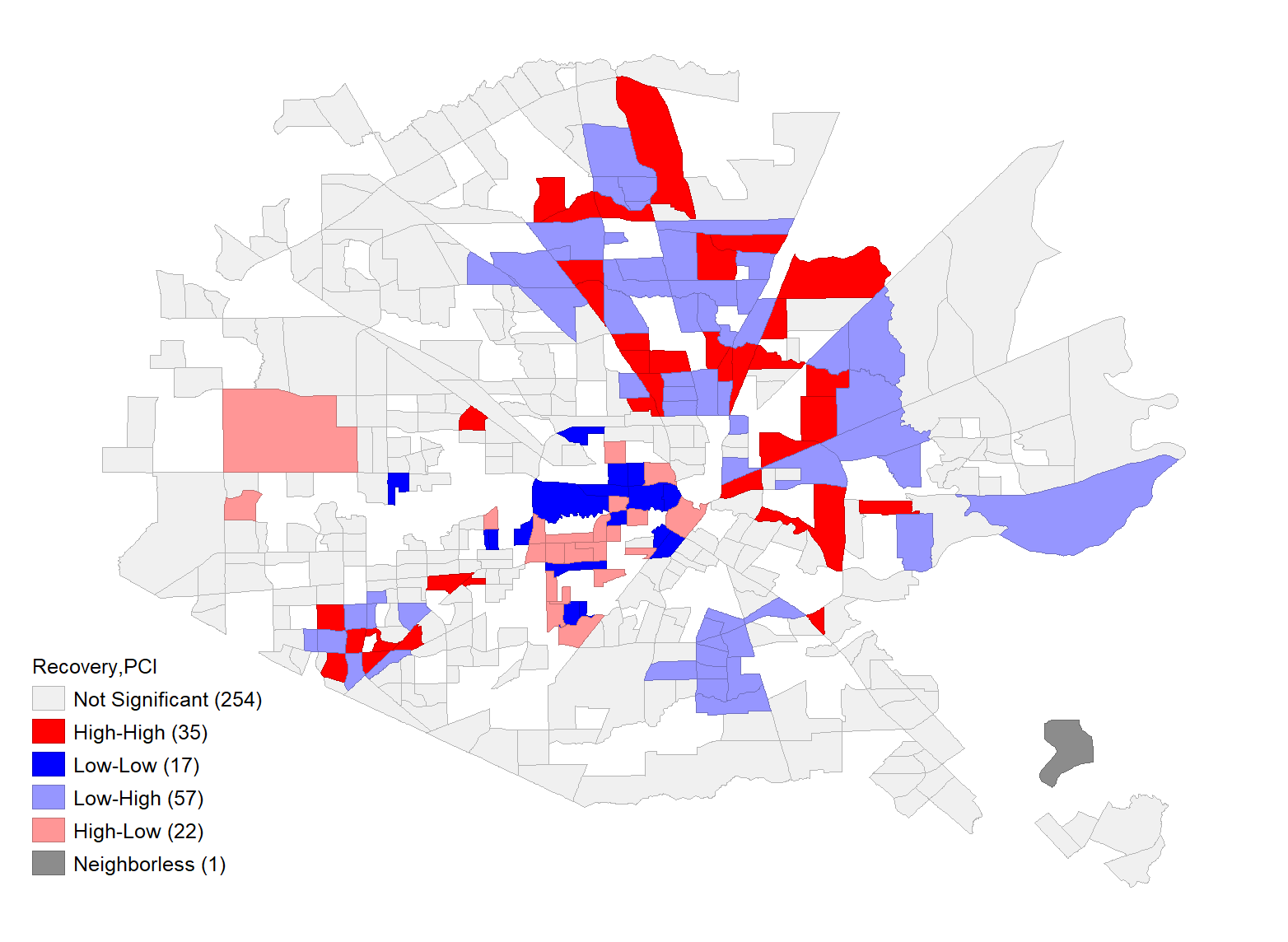}
\caption{LISA clustering of the duration of recovery for grocery stores to per capita income in Harris County. Recovery refers to the number of days of the disruption in grocery stores. PCI, or per capita income, refers to the income of the census tract divided by the population in the same census tract. High-high clusters refers to long recovery duration and high social vulnerability, or low PCI, and vice versa.}
\end{figure}

\section{Discussions and future directions}
This study proposed and demonstrated a smart flood resilience framework by harnessing heterogeneous community-scale big data and sensor data. We identified four essential components: (1) predictive flood risk mapping, (2) rapid impact assessment, (3) predictive infrastructure failure prediction and monitoring, and (4) smart situational awareness. This research sets up the first step to define the core capabilities of the smart flood resilience framework. Future research can focus on either new aspects of a capability (such as consideration of flood impact due to interactions among community residents and the built environment and infrastructure systems for the rapid impact assessment capability) or identification of new capabilities within the framework (such as the predictive compounding risk monitoring due to multiple disaster events). This study also suggests the development of new aspects of the proposed capabilities and new capabilities be driven by the practical needs of crisis managers. In addition, we introduced two examples of using various heterogeneous community-scale and sensor data in advanced data analytic approaches and machine learning techniques for demonstrating their capabilities within the smart flood resilience framework. Data categories and AI methods are not limited to the presented examples in this study. Future work in this field can concentrate on the investigation of additional datasets, such as satellite images and video camera data, and advanced AI methods, such as spatio-temporal graph neural networks, to augment the capabilities proposed in this study. We present a brief summary of emerging datasets and the vision of future AI methods for more comprehensive establishments and implementation of the smart flood resilience framework. 

The proposed smart flood resilience framework can also benefit implementations for practice for crisis response and recovery. Predictive flood risk and exposure mapping before hurricanes and floods can provide evacuation guidelines, such as which communities are more likely to become inundated and which roads have a high probability of flooding. Automated rapid impact assessment during floods can help crisis managers identify hotspots with community disruptions and more negative emotions and further deliver on-target relief resources. Predictive infrastructure failure predictions and monitoring during floods can help both crisis managers and impacted communities identify the roads to be flooded in 2, 4, or 6 hours, which can further benefit their route selections for evacuations and searching help from hospitals. Smart situational awareness during and after disasters can provide crisis managers with a better overview of regions with severe disaster impact and longer recovery durations than survey data, which can help them design recovery strategies by considering such spatial disparities. The field of smart flood resilience is expected to expand significantly. Two major drivers for the accelerated discovery and innovation in this field are growth in the availability and quality of emerging datasets and the advancements in the AI methods. In the final part of this paper, we discuss opportunities and challenges provided by these two major driving forces. 

\subsection{Opportunities and challenges in emerging data}
Community-scale big data, such as infrastructure sensor data, location-based population activity data, mobility data, and crowdsourced and social media data (Yabe et al. 2020; Schnebele et al. 2014), is becoming increasingly available through the growing use of sensing technologies, as well as “data for good” programs of commercial data aggregators and analytics companies. (Neelam and Sood 2020; Ianuale et al. 2015). Community-scale big data enables capture of the dynamics of community status, flooding evolutions, and the built environment (Eugene et al. 2021; Leitão et al. 2018). Accordingly, smart flood resilience can harness community-scale big data and develop appropriate analytic methods and machine learning approaches to augment its four core capabilities (e.g., Fan et al. 2020b; Podesta et al. 2021). This section summarizes 12 categories of emerging datasets that can support the implementation of core capabilities within smart flood resilience (Table 2). It is worth noting that emerging datasets for smart flood resilience were not limited to the summaries in Table 2.

To be specific, remote sensing data, such as satellite and drone images, can be used for rapid impact assessment and predictions of system recovery for various disaster types. Existing studies have used satellite images to evaluate and visualize post-hurricane damages using deep learning models (Cheng et al. 2021; McCarthy et al. 2020; Fujita et al. 2017). Using drone images, Alexandrov et al. (2019) established the machine learning models to detect wildfire smoke. There are also implementations of AI techniques with satellite image data for assessing building damages in earthquakes (Cooner et al. 2016). In addition, recent studies have employed satellite images for developing AI models to track changes in land cover and use for recovery assessment (Ghaffarian et al. 2021; Sheykhmousa et al. 2019). However, the use of such datasets depends heavily on deep learning techniques that bring challenges in the interpretation of model transparency and results explainability. Also, data quality of satellite images can be impacted by weather conditions, such as cloudy weather, can limit or even block the contents captured by satellite.

Crowdsourced data or volunteered geographic information (VGI) data can provide insights, through geospatial analytic approaches, into road inundations and the locations of people in need who made rescue requests. Hence, such insights can be further applied for training machine learning models to make predictions of road inundations, and for crisis managers to deliver on-target rescue support to people in need. Such datasets cannot cover data-scarce regions where few people use crowdsourced platforms to make rescue requests or regions with limited mobile signals so that people cannot use their smart devices to report their awareness of infrastructure failures. In addition, the reliability of these self-reported reports remains a critical issue, as well as the privacy issue of users reporting rescue requests as their home locations were visible on these platforms. Similarly, 3-1-1 call data cannot cover data-scarce areas and has limited representativeness of the general population; thus, 3-1-1 call data needs to be integrated with other datasets to augment rapid impact assessment capability. Despite these limitations, crowdsourced data provide useful additional insights that could inform flood impact assessment and situational awareness. In particular, crowdsourced data could provide insights in blind areas (areas where limited physical flood sensors exist). To fully leverage crowdsourced data, future studies should develop methods for evaluating the biases and reliability in crowdsourced data. Also, the integration of various crowdsourced and physical sensor data should be further investigated for improving the flood monitoring and situational awareness models.

\begin{landscape}

\begin{table}[]
\caption{Summary of emerging datasets for smart flood resilience assessments}
\footnotesize
\begin{tabular}{lllllllll}
\hline
Data Type                                                                     & \begin{tabular}[c]{@{}l@{}}Smart Resilience\\ Capability\end{tabular}                                                                                                     & Type of AI                                                                                                                               & \begin{tabular}[c]{@{}l@{}}Spatial\\ Resolution\end{tabular}    & \begin{tabular}[c]{@{}l@{}}Temporal\\ Resolution\end{tabular} & Time Lag       & \begin{tabular}[c]{@{}l@{}}Cost and\\ Availability\end{tabular}                     & \begin{tabular}[c]{@{}l@{}}Context (data\\ rich versus\\ data scarce)\end{tabular}    & Major Limitations                                                                                                \\ \hline
\begin{tabular}[c]{@{}l@{}}Satellite\\ Imagery\end{tabular}                   & \begin{tabular}[c]{@{}l@{}}Rapid Impact Assessment;\\ Predictive Monitoring of\\ Recovery\end{tabular}                                                                    & \begin{tabular}[c]{@{}l@{}}Computer vision;\\ Regression, etc.\end{tabular}                                                              & Coarse                                                          & \begin{tabular}[c]{@{}l@{}}Week or\\ Month\end{tabular}       & Days           & \begin{tabular}[c]{@{}l@{}}Higher resolution\\ images are \\ expensive\end{tabular} & \begin{tabular}[c]{@{}l@{}}Coverage of \\ data scarce\\ areas\end{tabular}            & \begin{tabular}[c]{@{}l@{}}Low resolution of \\ images for assessments\\ such as building\\ damages\end{tabular} \\
\begin{tabular}[c]{@{}l@{}}Crowdsourced\\ data\end{tabular}                   & \begin{tabular}[c]{@{}l@{}}Predictive Infrastructure\\ Failure Prediction and \\ Monitoring; Predictive\\ Monitoring of Response\end{tabular}                             & \begin{tabular}[c]{@{}l@{}}Geospatial analytic\\ approaches\end{tabular}                                                                 & Coordinate                                                      & \begin{tabular}[c]{@{}l@{}}Near \\ real-time\end{tabular}     & Hours, Minutes & \begin{tabular}[c]{@{}l@{}}Publicly\\ accessible\end{tabular}                       & \begin{tabular}[c]{@{}l@{}}Limited\\ coverage of \\ data scarce \\ areas\end{tabular} & \begin{tabular}[c]{@{}l@{}}Data reliability;\\ Data privacy issues\end{tabular}                                  \\
\begin{tabular}[c]{@{}l@{}}Social media \\ data\end{tabular}                  & \begin{tabular}[c]{@{}l@{}}Rapid Impact Assessment;\\ Predictive Infrastructure \\ Failure Prediction and\\ Monitoring; Predictive \\ Monitoring of Recovery\end{tabular} & \begin{tabular}[c]{@{}l@{}}Natural language\\ processing tools; \\ Computer vision; \\ Network analysis;\\ Machine learning\end{tabular} & \begin{tabular}[c]{@{}l@{}}Coordinate\\ or Polygon\end{tabular} & \begin{tabular}[c]{@{}l@{}}Near\\ real-time\end{tabular}      & Minutes        & \begin{tabular}[c]{@{}l@{}}Publicly \\ accessible\\ (limited)\end{tabular}          & \begin{tabular}[c]{@{}l@{}}Limited\\ coverage of\\ data scarce \\ areas\end{tabular}  & \begin{tabular}[c]{@{}l@{}}Data reliability and\\ representativeness\end{tabular}                                \\
\begin{tabular}[c]{@{}l@{}}Telemetry calls\\ data\end{tabular}                & Rapid Impact Assessment                                                                                                                                                   & \begin{tabular}[c]{@{}l@{}}Geospatial analytic\\ approaches;\\ Statistical techniques\end{tabular}                                       & Coordinate                                                      & \begin{tabular}[c]{@{}l@{}}Near\\ real-time\end{tabular}      & Hours, Minutes & \begin{tabular}[c]{@{}l@{}}Publicly\\ accessible\\ (limited regions)\end{tabular}   & \begin{tabular}[c]{@{}l@{}}Limited\\ coverage of\\ data scarce \\ areas\end{tabular}  & \begin{tabular}[c]{@{}l@{}}Data\\ representativeness\end{tabular}                                                \\
\begin{tabular}[c]{@{}l@{}}Population activity \\ telemetry data\end{tabular} & Rapid Impact Assessment                                                                                                                                                   & \begin{tabular}[c]{@{}l@{}}Geospatial analytic\\ approaches; \\ Machine learning\end{tabular}                                            & \begin{tabular}[c]{@{}l@{}}Polygon\\ (region)\end{tabular}      & Four-hour                                                     & Weeks          & \begin{tabular}[c]{@{}l@{}}Publicly\\ accessible\\ (at request)\end{tabular}        & \begin{tabular}[c]{@{}l@{}}Coverage of\\ data scarce\\ areas\end{tabular}             & \begin{tabular}[c]{@{}l@{}}Relative activity\\ index\end{tabular}                                                \\
Streetlight data                                                              & \begin{tabular}[c]{@{}l@{}}Predictive Monitoring of\\ Recovery\end{tabular}                                                                                               & \begin{tabular}[c]{@{}l@{}}Graph mining;\\ Network analysis\end{tabular}                                                                 & \begin{tabular}[c]{@{}l@{}}Polygon\\ (region)\end{tabular}      & Hourly                                                        & Hours          & \begin{tabular}[c]{@{}l@{}}Publicly\\ accessible\\ (at request)\end{tabular}        & \begin{tabular}[c]{@{}l@{}}Coverage of \\ data scarce\\ areas\end{tabular}            & \begin{tabular}[c]{@{}l@{}}Low spatial\\ resolution\end{tabular}                                                 \\
\begin{tabular}[c]{@{}l@{}}Population\\ mobility data\end{tabular}            & Rapid Impact Assessment                                                                                                                                                   & \begin{tabular}[c]{@{}l@{}}Graph neural networks;\\ Reinforcement learning;\\ Long short-term\\ memory\end{tabular}                      & Coordinate                                                      & \begin{tabular}[c]{@{}l@{}}Near\\ real-time\end{tabular}      & Days           & \begin{tabular}[c]{@{}l@{}}Limited access\\ with high \\ purchase cost\end{tabular} & \begin{tabular}[c]{@{}l@{}}Coverage of\\ data scarce\\ areas\end{tabular}             & \begin{tabular}[c]{@{}l@{}}Noise and\\ uncertainty due to\\ the pattern of \\ mobile devices\end{tabular}        \\
\begin{tabular}[c]{@{}l@{}}Mobile phone\\ activity data\end{tabular}          & \begin{tabular}[c]{@{}l@{}}Predictive Monitoring of\\ Recovery\end{tabular}                                                                                               & \begin{tabular}[c]{@{}l@{}}Geospatial analytic\\ approaches\end{tabular}                                                                 & \begin{tabular}[c]{@{}l@{}}Polygon\\ (region)\end{tabular}      & Hourly                                                        & Weeks, Months  & \begin{tabular}[c]{@{}l@{}}Publicly \\ accessible\\ (at request)\end{tabular}       & \begin{tabular}[c]{@{}l@{}}Coverage of\\ data scarce\\ areas\end{tabular}             & \begin{tabular}[c]{@{}l@{}}Data\\ representativeness\end{tabular}                                                \\
\begin{tabular}[c]{@{}l@{}}Credit card\\ transactions\end{tabular}            & \begin{tabular}[c]{@{}l@{}}Predictive Monitoring of\\ Recovery\end{tabular}                                                                                               & \begin{tabular}[c]{@{}l@{}}Geospatial analytic\\ approaches\end{tabular}                                                                 & \begin{tabular}[c]{@{}l@{}}Polygon\\ (region)\end{tabular}      & Daily                                                         & Weeks, Months  & \begin{tabular}[c]{@{}l@{}}Publicly\\ accessible\\ (at request)\end{tabular}        & \begin{tabular}[c]{@{}l@{}}Coverage of\\ data scarce\\ areas\end{tabular}             & \begin{tabular}[c]{@{}l@{}}Data\\ representativeness\end{tabular}                                                \\
Traffic data                                                                  & \begin{tabular}[c]{@{}l@{}}Predictive Infrastructure\\ Failure Prediction and \\ Monitoring\end{tabular}                                                                  & \begin{tabular}[c]{@{}l@{}}Spatio-temporal Graphic\\ Neural Network;\\ Deep learning;\\ Contagion model\end{tabular}                     & Coordinate                                                      & \begin{tabular}[c]{@{}l@{}}Five-\\ minute\end{tabular}        & Days           & \begin{tabular}[c]{@{}l@{}}Limited access\\ with high\\ purchase cost\end{tabular}  & \begin{tabular}[c]{@{}l@{}}Coverage of\\ data scarce\\ areas\end{tabular}             & \begin{tabular}[c]{@{}l@{}}Large data size\\ requiring significant\\ computation efforts\end{tabular}            \\
\begin{tabular}[c]{@{}l@{}}Traffic\\ camera data\end{tabular}                 & \begin{tabular}[c]{@{}l@{}}Predictive Infrastructure\\ Failure Prediction and\\ Monitoring\end{tabular}                                                                   & \begin{tabular}[c]{@{}l@{}}Bayesian Network;\\ Computer vision\end{tabular}                                                              & Coordinate                                                      & Real-time                                                     & Minutes        & \begin{tabular}[c]{@{}l@{}}High cost for\\ many cameras\end{tabular}                & \begin{tabular}[c]{@{}l@{}}No coverage\\ of data scarce\\ areas\end{tabular}          & \begin{tabular}[c]{@{}l@{}}Limited deployment\\ of sensors; Significant\\ computation efforts\end{tabular}       \\
\begin{tabular}[c]{@{}l@{}}Flood\\ sensor data\end{tabular}                   & \begin{tabular}[c]{@{}l@{}}Predictive Flood Risk\\ Mapping; Rapid Impact \\ Assessment; Predictive \\ Infrastructure Failure \\ Prediction and Monitoring\end{tabular}    & \begin{tabular}[c]{@{}l@{}}Deep learning;\\ Bayesian modelling\end{tabular}                                                              & Coordinate                                                      & Hourly                                                        & One hour       & \begin{tabular}[c]{@{}l@{}}Publicly\\ accessible\end{tabular}                       & \begin{tabular}[c]{@{}l@{}}No coverage\\ of data scarce\\ areas\end{tabular}          & \begin{tabular}[c]{@{}l@{}}Limited deployment\\ of sensors\end{tabular}                                          \\ \hline
\end{tabular}
\end{table}
\end{landscape}

Social media data (especially Twitter data) has been employed for rapid impact assessment, predictive infrastructure failure predictions, and monitoring of response and recovery using natural language processing tools (e.g., sentiment analysis and topic modeling), computer vision, and machine learning techniques. These datasets were previously available at coordinate scale, but now are available only at polygon scale now due to Twitter policies. Although social media data can be near real-time, it cannot cover data scarce areas. It also has limited representativeness of the general population, and its reliability remains in doubt. Hence, to more effectively use social media data for situational awareness in flood events, future studies should focus on methods for de-biasing datasets and algorithms to improve fairness in disaster informatics tasks conducted on social media data (Yang et al. 2020).

Location-based human activity datasets can reveal the patterns of human movements to assess flood impact and community recovery through graph mining, machine learning, and network analysis. Telemetry-based cellphone activity data could be used to yield a relative activity index within a spatial area (e.g., census tract, ZIP code, and census block group). The fluctuations in the activity index could provide signals about flood impacts and the extent of recovery. GPS-based human mobility data can provide insights regarding the mobility activity across different spatial areas. The analysis of human mobility patterns can provide insights regarding the preparedness, evacuation, response, and recovery of different areas. Location-based data can also shed light on the human activities, such as evaluation of visits to POIs and their fluctuations, which could provide useful insights regarding flood impacts and recovery. Location-based data have two major limitations. First is their high cost. These datasets are usually provided by third-party companies. Due to cost, their use could be limited in low-resource communities. Another limitation is related to potential biases in location-based data. Not all location-based datasets provide a representative sample for the population, and hence, their use in impact and recovery assessment could lead to biased insights. To address this limitation, future studies should consider evaluating biases in mobility datasets and develop de-biasing methods to enhance the datasets before they are used in flood impact and recovery assessment studies.

As shown in Table 2, several data types could be leveraged for augmenting smart flood resilience. With advancements in technologies and research, the resolution and availability of such data are expected to improve in the years to come. These expected advancements provide opportunities to harness these heterogeneous datasets in creating and testing proper machine learning and deep-learning models for various smart flood resilience capabilities. In fact, the advancements in the field of AI and machine learning could be another driver to the field of smart flood resilience. We discuss examples of such advancements in the next sub-section.

\subsection{Advancements in AI and computing techniques}
Recent and future advances in the field of AI and machine learning could provide the much needed computational models and algorithms to augment smart flood resilience. In this section, we elaborate upon examples of methods in the frontier of machine learning and deep learning that could be leveraged in the field of smart resilience. First, the spread of floodwaters in urban networks is a complex phenomenon that requires models and tools for predicting both spatial and temporal patterns of flood risk. Recent advances in deep learning research (Zhang et al. 2019; Yu et al. 2017) have highlighted the powerful capability of spatio-temporal graph learning models. These models push the boundary of existing variants of graph representation models, such as graph convolutional network, graph attention network and graph recurrent network, through encoding the spatial and temporal information of the nodes (Yan et al. 2018). Specifically, the spatiotemporal graph convolutional network has been proposed to satisfy the requirements of mid- and long-term prediction tasks and to tackle the time series prediction by considering the spatial and temporal dependencies (Yuan et al. 2021b). Second, the management success and efficiency of the flooding response system depends heavily on the ability to obtain, assess, and communicate information in a timely manner. Such information is collected from multiple information channels and sources (e.g., Eugene et al. 2021, Fan and Mostafavi 2019). Traditional data processing approaches require centralizing the training data on one machine or in a data center. This process delays data set collection and prediction result delivery. Recent advances in machine learning research have proposed federated learning, which allows individual models to collaboratively learn a shared prediction model while keeping all training data on the device (Bonawitz et al. 2019). This machine learning technique decouples the ability to effect machine learning from the need to store the data in the cloud and brings model training to the device as well (Li et al. 2020). Flood response will significantly benefit from the advancements of the federated learning technology, as it addresses the need for data sharing among organizations. Organizations can have models trained on their data (without the need for sharing the data with other organizations), but their models collaborate in a shared flood risk prediction task.

Third, during the data sensing process, collected multi-modal data may have potential fairness issues for model training, which could potentially lead to unfair monitoring results regarding flood status. Recent studies in machine learning have developed novel approaches to mitigate fairness issues. For example, the importance sampling strategy for the multi-modal data collection, in which data samples are assigned with adaptive weights based on their sensitive attributes (e.g., race and income of spatial areas) and these standards are widely adopted to mitigate the biases in data samples (Wang et al. 2018). Specifically, we could assign equal importance to each unique sample at the beginning and gradually update the importance matrix along with the data sampling process (Cappé et al. 2008). The updating rule reduces the importance score for those data samples containing existing attribute values to lower their importance and contributions for future system training. In this way, higher priorities would be given to those samples with sensitive attributes (e.g., race and income), thus achieving better fairness regarding the model predictions for flood status (O'Reilly-Shah et al. 2020). As such, fair models can be applied to various data types and prediction tasks, including the spatial-temporal data in our flood monitoring task. This would contribute to smart and fair disaster response and recovery during future flooding events.


\section*{Acknowledgements}
The authors would like to acknowledge funding support from the National Science Foundation CRISP 2.0 Type 2 \#1832662 and the X-Grant program (Presidential Excellence Fund) from Texas A\&M University. The authors would also like to acknowledge INRIX, Inc. and SafeGraph for providing data. Any opinions, findings, and conclusions or recommendations expressed in this research are those of the authors and do not necessarily reflect the views of the funding agencies.

\section*{Data availability}
The data that support the findings of this study are available fromINRIX, Inc. and SafeGraph, but restrictions apply to the availability of these data, which were used under license for the current study. The data can be accessed upon request submitted on each data provider. Other data (flood inundations and flood claims) we use in this study are all publicly available.

\section*{Code availability}
The code that supports the findings of this study is available from the corresponding author upon request.

\section*{References}
Adam, N. R., Shafiq, B., \& Staffin, R. (2012). Spatial computing and social media in the context of disaster management. IEEE Intelligent Systems, 27(6), 90-96.

Alazawi, Z., Alani, O., Abdljabar, M. B., Altowaijri, S., \& Mehmood, R. (2014). A smart disaster management system for future cities. In Proceedings of the 2014 ACM international workshop on wireless and mobile technologies for smart cities (pp. 1-10).

Alexandrov, D., Pertseva, E., Berman, I., Pantiukhin, I., \& Kapitonov, A. (2019). Analysis of machine learning methods for wildfire security monitoring with an unmanned aerial vehicles. In 2019 24th conference of open innovations association (FRUCT) (pp. 3-9). IEEE.

Anbarasan, M., Muthu, B., Sivaparthipan, C. B., Sundarasekar, R., Kadry, S., Krishnamoorthy, S., \& Dasel, A. A. (2020). Detection of flood disaster system based on IoT, big data and convolutional deep neural network. Computer Communications, 150, 150-157.

Beatty, T. K., Shimshack, J. P., \& Volpe, R. J. (2019). Disaster preparedness and disaster response: Evidence from sales of emergency supplies before and after hurricanes. Journal of the Association of Environmental and Resource Economists, 6(4), 633-668.

Bonawitz, K., Eichner, H., Grieskamp, W., Huba, D., Ingerman, A., Ivanov, V., ... \& Roselander, J. (2019). Towards federated learning at scale: System design. arXiv preprint arXiv:1902.01046.Boon, H. J. (2014). Disaster resilience in a flood-impacted rural Australian town. Natural hazards, 71(1), 683-701.

Boukerche, A. (2019). Smart Disaster Management and Responses for Smart Cities: A new Challenge for the Next Generation of Distributed Simulation Systems. In 2019 IEEE/ACM 23rd International Symposium on Distributed Simulation and Real Time Applications (DS-RT) (pp. 1-2). IEEE.

Cameron, M. A., Power, R., Robinson, B., \& Yin, J. (2012). Emergency situation awareness from twitter for crisis management. In Proceedings of the 21st international conference on world wide web (pp. 695-698).

Cappé, O., Douc, R., Guillin, A., Marin, J. M., \& Robert, C. P. (2008). Adaptive importance sampling in general mixture classes. Statistics and Computing, 18(4), 447-459.

Cervone, G., Sava, E., Huang, Q., Schnebele, E., Harrison, J., \& Waters, N. (2016). Using Twitter for tasking remote-sensing data collection and damage assessment: 2013 Boulder flood case study. International Journal of Remote Sensing, 37(1), 100-124.

Cheng, C. S., Behzadan, A. H., \& Noshadravan, A. (2021). Deep learning for post‐hurricane aerial damage assessment of buildings. Computer‐Aided Civil and Infrastructure Engineering.

Cooner, A. J., Shao, Y., \& Campbell, J. B. (2016). Detection of urban damage using remote sensing and machine learning algorithms: Revisiting the 2010 Haiti earthquake. Remote Sensing, 8(10), 868.

Dong, S., Yu, T., Farahmand, H., \& Mostafavi, A. (2020a). Bayesian modeling of flood control networks for failure cascade characterization and vulnerability assessment. Computer‐Aided Civil and Infrastructure Engineering, 35(7), 668-684.

Dong, S., Yu, T., Farahmand, H., \& Mostafavi, A. (2020b). Probabilistic modeling of cascading failure risk in interdependent channel and road networks in urban flooding. Sustainable Cities and Society, 62, 102398.

Dong, S., Esmalian, A., Farahmand, H., \& Mostafavi, A. (2020c). An integrated physical-social analysis of disrupted access to critical facilities and community service-loss tolerance in urban flooding. Computers, Environment and Urban Systems, 80, 101443.

Esmalian, A., Dong, S., Coleman, N., \& Mostafavi, A. (2019). Determinants of risk disparity due to infrastructure service losses in disasters: a household service gap model. Risk analysis.

Eugene, A., Alpert, N., Lieberman-Cribbin, W., \& Taioli, E. (2021). Using NYC 311 Call Center Data to Assess Short-and Long-Term Needs Following Hurricane Sandy. Disaster Medicine and Public Health Preparedness, 1-5.

Fan, C., Wu, F., \& Mostafavi, A. (2020a). A hybrid machine-learning pipeline for automated mapping of events and locations from social media in disasters. IEEE Access, 8, 10478-10490.

Fan, C., Jiang, X., \& Mostafavi, A. (2020b). A network percolation-based contagion model of flood propagation and recession in urban road networks. Scientific Reports, 10(1), 1-12.

Fan, C., \& Mostafavi, A. (2019). A graph‐based method for social sensing of infrastructure disruptions in disasters. Computer‐Aided Civil and Infrastructure Engineering, 34(12), 1055-1070.

Fujita, A., Sakurada, K., Imaizumi, T., Ito, R., Hikosaka, S., \& Nakamura, R. (2017, May). Damage detection from aerial images via convolutional neural networks. In 2017 Fifteenth IAPR international conference on machine vision applications (MVA) (pp. 5-8). IEEE.

Ghaffarian, S., Roy, D., Filatova, T., \& Kerle, N. (2021). Agent-based modelling of post-disaster recovery with remote sensing data. International Journal of Disaster Risk Reduction, 60, 102285.

Harris County Flood Control District. (2021). Harris County Has Never Seen a Storm Like Harvey. https://www.hcfcd.org/About/Harris-Countys-Flooding-History/Hurricane-Harvey (accessed Oct 4, 2021).

Helderop, E., \& Grubesic, T. H. (2019). Flood evacuation and rescue: The identification of critical road segments using whole-landscape features. Transportation research interdisciplinary perspectives, 3, 100022.

Hikichi, H., Tsuboya, T., Aida, J., Matsuyama, Y., Kondo, K., Subramanian, S. V., \& Kawachi, I. (2017). Social capital and cognitive decline in the aftermath of a natural disaster: a natural experiment from the 2011 Great East Japan Earthquake and Tsunami. The Lancet Planetary Health, 1(3), e105-e113.

Huang, Q., \& Xiao, Y. (2015). Geographic situational awareness: mining tweets for disaster preparedness, emergency response, impact, and recovery. ISPRS International Journal of Geo-Information, 4(3), 1549-1568.

Hughes, D., Greenwood, P., Blair, G., Coulson, G., Pappenberger, F., Smith, P., \& Beven, K. (2006). An intelligent and adaptable grid-based flood monitoring and warning system. In Proceedings of the UK eScience All Hands Meeting (Vol. 10).

Ianuale, N., Schiavon, D., \& Capobianco, E. (2015). Smart cities, big data, and communities: Reasoning from the viewpoint of attractors. IEEE Access, 4, 41-47.

Leitão, J. P., Peña-Haro, S., Lüthi, B., Scheidegger, A., \& de Vitry, M. M. (2018). Urban overland runoff velocity measurement with consumer-grade surveillance cameras and surface structure image velocimetry. Journal of Hydrology, 565, 791-804.

Li, T., Sahu, A. K., Talwalkar, A., \& Smith, V. (2020). Federated learning: Challenges, methods, and future directions. IEEE Signal Processing Magazine, 37(3), 50-60.

Li, Z., Wang, C., Emrich, C. T., \& Guo, D. (2018). A novel approach to leveraging social media for rapid flood mapping: a case study of the 2015 South Carolina floods. Cartography and Geographic Information Science, 45(2), 97-110.

Lu, X., Wrathall, D. J., Sundsøy, P. R., Nadiruzzaman, M., Wetter, E., Iqbal, A., ... \& Bengtsson, L. (2016). Unveiling hidden migration and mobility patterns in climate stressed regions: A longitudinal study of six million anonymous mobile phone users in Bangladesh. Global Environmental Change, 38, 1-7.

MacEachren, A. M., Jaiswal, A., Robinson, A. C., Pezanowski, S., Savelyev, A., Mitra, P., ... \& Blanford, J. (2011). Senseplace2: Geotwitter analytics support for situational awareness. In 2011 IEEE conference on visual analytics science and technology (VAST) (pp. 181-190). IEEE.

McCarthy, M. J., Jessen, B., Barry, M. J., Figueroa, M., McIntosh, J., Murray, T., ... \& Muller-Karger, F. E. (2020). Mapping hurricane damage: A comparative analysis of satellite monitoring methods. International Journal of Applied Earth Observation and Geoinformation, 91, 102134.

Meier, P. (2015). Digital humanitarians: how big data is changing the face of humanitarian response. Crc Press.

Morss, R. E., Mulder, K. J., Lazo, J. K., \& Demuth, J. L. (2016). How do people perceive, understand, and anticipate responding to flash flood risks and warnings? Results from a public survey in Boulder, Colorado, USA. Journal of hydrology, 541, 649-664.

Muhammad, A. N., Aseere, A. M., Chiroma, H., Shah, H., Gital, A. Y., \& Hashem, I. A. T. (2020). Deep learning application in smart cities: recent development, taxonomy, challenges and research prospects. Neural Computing and Applications, 1-37.

Nan, C., \& Sansavini, G. (2017). A quantitative method for assessing resilience of interdependent infrastructures. Reliability Engineering \& System Safety, 157, 35-53.

Naulin, J. P., Payrastre, O., \& Gaume, E. (2013). Spatially distributed flood forecasting in flash flood prone areas: Application to road network supervision in Southern France. Journal of Hydrology, 486, 88-99.

Neelam, S., \& Sood, S. K. (2020). A scientometric review of global research on smart disaster management. IEEE Transactions on Engineering Management, 68(1), 317-329.

O'Reilly-Shah, V. N., Gentry, K. R., Walters, A. M., Zivot, J., Anderson, C. T., \& Tighe, P. J. (2020). Bias and ethical considerations in machine learning and the automation of perioperative risk assessment. British Journal of Anaesthesia, 125(6), 843-846.

Palen, L., \& Anderson, K. M. (2016). Crisis informatics—New data for extraordinary times. Science, 353(6296), 224-225.

Plank, S. (2014). Rapid damage assessment by means of multi-temporal SAR—A comprehensive review and outlook to Sentinel-1. Remote Sensing, 6(6), 4870-4906.

Podesta, C., Coleman, N., Esmalian, A., Yuan, F., \& Mostafavi, A. (2021). Quantifying community resilience based on fluctuations in visits to points-of-interest derived from digital trace data. Journal of the Royal Society Interface, 18(177), 20210158.

Praharaj, S., Chen, T. D., Zahura, F. T., Behl, M., \& Goodall, J. L. (2021). Estimating impacts of recurring flooding on roadway networks: a Norfolk, Virginia case study. Natural Hazards, 1-25.

Roy, K. C., Hasan, S., \& Mozumder, P. (2020). A multilabel classification approach to identify hurricane‐induced infrastructure disruptions using social media data. Computer‐Aided Civil and Infrastructure Engineering, 35(12), 1387-1402.

Schnebele, E., Cervone, G., \& Waters, N. (2014). Road assessment after flood events using non-authoritative data. Natural Hazards and Earth System Sciences, 14(4), 1007-1015.

See, L. (2019). A review of citizen science and crowdsourcing in applications of pluvial flooding. Frontiers in Earth Science, 7, 44.

Sherrieb, K., Norris, F. H., \& Galea, S. (2010). Measuring capacities for community resilience. Social indicators research, 99(2), 227-247.

Sheykhmousa, M., Kerle, N., Kuffer, M., \& Ghaffarian, S. (2019). Post-disaster recovery assessment with machine learning-derived land cover and land use information. Remote sensing, 11(10), 1174.

Silver, A., \& Matthews, L. (2017). The use of Facebook for information seeking, decision support, and self-organization following a significant disaster. Information, Communication \& Society, 20(11), 1680-1697.

Sood, S. K., Sandhu, R., Singla, K., \& Chang, V. (2018). IoT, big data and HPC based smart flood management framework. Sustainable Computing: Informatics and Systems, 20, 102-117.

Sun, W., Bocchini, P., \& Davison, B. D. (2020). Applications of artificial intelligence for disaster management. Natural Hazards, 1-59.
United Nations Office for Disaster Risk Reduction (UNDRR). (2015). Sendai Framework for Disaster Risk Reduction 2015-2030. https://www.preventionweb.net/files/43291\_sendaiframeworkfordrren.pdf (accessed Oct 4, 2021)

Vieweg, S., Hughes, A. L., Starbird, K., \& Palen, L. (2010). Microblogging during two natural hazards events: what twitter may contribute to situational awareness. In Proceedings of the SIGCHI conference on human factors in computing systems (pp. 1079-1088).

Vugrin, E. D., Warren, D. E., Ehlen, M. A., \& Camphouse, R. C. (2010). A framework for assessing the resilience of infrastructure and economic systems. In Sustainable and resilient critical infrastructure systems (pp. 77-116). Springer, Berlin, Heidelberg.

Wang, Q., Phillips, N. E., Small, M. L., \& Sampson, R. J. (2018). Urban mobility and neighborhood isolation in America’s 50 largest cities. Proceedings of the National Academy of Sciences, 115(30), 7735-7740.

Wang, W., Yang, S., Stanley, H. E., \& Gao, J. (2019). Local floods induce large-scale abrupt failures of road networks. Nature communications, 10(1), 1-11.

Wing, O. E., Pinter, N., Bates, P. D., \& Kousky, C. (2020). New insights into US flood vulnerability revealed from flood insurance big data. Nature communications, 11(1), 1-10.

Yabe, T., Tsubouchi, K., Fujiwara, N., Sekimoto, Y., \& Ukkusuri, S. V. (2020). Understanding post-disaster population recovery patterns. Journal of the Royal Society Interface, 17(163), 20190532.

Yan, S., Xiong, Y., \& Lin, D. (2018). Spatial temporal graph convolutional networks for skeleton-based action recognition. In Thirty-second AAAI conference on artificial intelligence.

Yang, Y., Zhang, C., Fan, C., Mostafavi, A., \& Hu, X. (2020). Towards Fairness-Aware Disaster Informatics: an Interdisciplinary Perspective. IEEE Access, 8, 201040-201054.

Yin, J., Yu, D., \& Wilby, R. (2016a). Modelling the impact of land subsidence on urban pluvial flooding: A case study of downtown Shanghai, China. Science of the Total Environment, 544, 744-753.

Yin, J., Yu, D., Yin, Z., Liu, M., \& He, Q. (2016b). Evaluating the impact and risk of pluvial flash flood on intra-urban road network: A case study in the city center of Shanghai, China. Journal of hydrology, 537, 138-145.

Yu, B., Yin, H., \& Zhu, Z. (2017). Spatio-temporal graph convolutional networks: A deep learning framework for traffic forecasting. arXiv preprint arXiv:1709.04875.

Yuan, F., \& Liu, R. (2020). Mining social media data for rapid damage assessment during Hurricane Matthew: Feasibility study. Journal of Computing in Civil Engineering, 34(3), 05020001.

Yuan, F., \& Liu, R. (2018). Feasibility study of using crowdsourcing to identify critical affected areas for rapid damage assessment: Hurricane Matthew case study. International journal of disaster risk reduction, 28, 758-767.

Yuan, F., Liu, R., Mao, L., \& Li, M. (2021a). Internet of people enabled framework for evaluating performance loss and resilience of urban critical infrastructures. Safety science, 134, 105079.

Yuan, F., Xu, Y., Li, Q., \& Mostafavi, A. (2021b). Spatio-Temporal Graph Convolutional Networks for Road Network Status Prediction in Floods. arXiv preprint arXiv:2104.02276.

Yuan, F., Esmalian, A., Oztekin, B., \& Mostafavi, A. (2021c). Unveiling Spatial Patterns of Disaster Impacts and Recovery Using Credit Card Transaction Variances. arXiv preprint arXiv:2101.10090.

Yuan, F., Li, M., Liu, R., Zhai, W., \& Qi, B. (2021d). Social media for enhanced understanding of disaster resilience during Hurricane Florence. International Journal of Information Management, 57, 102289.

Zhai, W., Peng, Z. R., \& Yuan, F. (2020). Examine the effects of neighborhood equity on disaster situational awareness: Harness machine learning and geotagged Twitter data. International Journal of Disaster Risk Reduction, 48, 101611.

Zhang, Y., Cheng, T., \& Ren, Y. (2019). A graph deep learning method for short‐term traffic forecasting on large road networks. Computer‐Aided Civil and Infrastructure Engineering, 34(10), 877-896.

Zhang, C., Yao, W., Yang, Y., Huang, R., \& Mostafavi, A. (2020). Semiautomated social media analytics for sensing societal impacts due to community disruptions during disasters. Computer‐Aided Civil and Infrastructure Engineering, 35(12), 1331-1348.

\end{document}